\documentclass[iop]{emulateapj}
\usepackage{color}
\usepackage{graphicx}
\usepackage{natbib}
\usepackage{hyperref}
\usepackage{lmodern}
\usepackage{chngcntr}
\hypersetup{
  colorlinks   = true, 
  urlcolor     = blue,  
  linkcolor    = blue, 
  citecolor   = blue }    
\begin{document}

\title{Investigating Trends in Atmospheric Compositions of Cool Gas Giant Planets Using \emph{Spitzer} Secondary Eclipses}
\shorttitle{Trends in Atmospheric Compositions from \emph{Spitzer} Secondary Eclipses}
\shortauthors{Wallack et al.}
\author{Nicole L. Wallack$^{1}$, Heather A. Knutson$^{1}$, Caroline V. Morley$^{2,3}$, Julianne I. Moses$^{4}$, Nancy H. Thomas$^{1}$, Daniel P. Thorngren$^{5}$, Drake Deming$^{6}$, Jean-Michel D{\'e}sert$^{7}$, Jonathan J. Fortney$^{8}$, and Joshua A. Kammer$^{9}$}

\affil{$^{1}$Division of Geological \& Planetary Sciences, California Institute of Technology, Pasadena, CA 91125, USA \\
$^{2}$Department of Astronomy, Harvard University, Cambridge, MA 02138, USA \\
$^{3}$Department of Astronomy, University of Texas at Austin, Austin, TX 78712, USA \\
$^{4}$Space Science Institute, 4750 Walnut Street, Suite 205, Boulder, CO 80301, USA \\ 
$^{5}$Department of Physics, University of California, Santa Cruz, CA 95064, USA \\
$^{6}$Department of Astronomy, University of Maryland at College Park, College Park, MD 20742, USA\\
$^{7}$Anton Pannekoek Institute for Astronomy, University of Amsterdam, 1090 GE Amsterdam, The Netherlands \\
$^{8}$Department of Astronomy and Astrophysics, University of California, Santa Cruz, CA 95064, USA \\
$^{9}$Southwest Research Institute, San Antonio, TX 78238, USA \\}

\email{nwallack@caltech.edu}

\begin{abstract}
We present new 3.6 and 4.5~$\mu$m secondary eclipse measurements for five cool (T $\lesssim$ 1000 K) transiting gas giant planets: HAT-P-15b, HAT-P-17b, HAT-P-18b, HAT-P-26b, and WASP-69b. We detect eclipses in at least one bandpass for all planets except HAT-P-15b.  We confirm and refine the orbital eccentricity of HAT-P-17b, which is also the only planet in our sample with a known outer companion. We compare our measured eclipse depths in these two bands, which are sensitive to the relative abundances of methane versus carbon monoxide and carbon dioxide, respectively, to predictions from 1D atmosphere models for each planet.  For planets with hydrogen-dominated atmospheres and equilibrium temperatures cooler than $\sim$1000 K, this ratio should vary as a function of both atmospheric metallicity and the carbon-to-oxygen ratio. For HAT-P-26b, our observations are in good agreement with the low atmospheric metallicity inferred from transmission spectroscopy.  We find that all four of the planets with detected eclipses are best matched by models with relatively efficient circulation of energy to the nightside. We see no evidence for a solar-system-like correlation between planet mass and atmospheric metallicity, but instead identify a potential ($1.9\sigma$) correlation between the inferred CH$_4$/(CO+CO$_2$) ratio and stellar metallicity. Our ability to characterize this potential trend is limited by the relatively large uncertainties in the stellar metallicity values. Our observations provide a first look at the brightness of these planets at wavelengths accessible to the \textit{James Webb Space Telescope}, which will be able to resolve individual CH$_4$, CO, and CO$_2$ bands and provide much stronger constraints on their atmospheric compositions.
\end{abstract}

\section{Introduction}
Observations of the ever-expanding ensemble of exoplanetary systems provide unique statistical insights into the formation and evolution of planetary systems.  This is perhaps best illustrated by the classic correlation between gas giant planet frequency and host star metallicity \citep{Fischer2005TheCorrelation}, which suggests that these planets most likely formed via core accretion (e.g., \citealt{Pollack1996FormationGas}; \citealt{Johansen2017FormingAccretion}).
Observations of the masses and radii of extrasolar gas giant planets also indicate that, like the giant planets in our solar system, the average densities of these planets tend to increase with decreasing mass (\citealt{Miller2011TheRevealed}; \citealt{Thorngren2016ThePlanets}).  These trends are consistent with a picture in which Jovian-mass planets were able to accrete substantially more gas from the protoplanetary disk than Neptune-mass planets, either because their cores reached the critical mass for gas accretion earlier or because they formed in a region of the disk with a higher gas surface density (e.g., at shorter orbital periods).

The atmospheric compositions of gas giant planets should in theory allow us to distinguish between these two scenarios, as the incorporation of solids into the growing planet's atmosphere will enrich its bulk metallicity and leave a unique compositional fingerprint that will vary according to its formation location and epoch (e.g., \citealt{Oberg2011TheAtmospheres}; \citealt{Espinoza2017MetalExoplanets}). In the solar system, Jupiter has both a smaller core mass fraction and a lower atmospheric carbon-to-hydrogen ratio than Neptune (e.g., \citealt{Lodders2003SOLARLodders}).  However, with only one planetary system it is difficult to determine the relative importance of the formation location in determining atmospheric metallicity.  If we instead consider the broader population of exo-Jupiters and exo-Neptunes, which presumably originate from a variety of formation locations, we can ask whether exo-Neptunes consistently exhibit higher atmospheric metallicities than exo-Jupiters (and therefore whether such enhancement is largely independent of formation location) or whether both populations span a wide range of atmospheric metallicities that reflect their varied formation locations and accretion histories \citep{Humphries2018ChangesAccretion}.

\begin{centering}
\begin{deluxetable*}{lccccc}[!t]
\tablecaption{System Properties}
\tablewidth{0pt}
\tablehead{
\colhead{}&\colhead{HAT-P-15b}&\colhead{HAT-P-17b} &\colhead{HAT-P-18b}&\colhead{HAT-P-26b}&\colhead{WASP-69b}}
\startdata
T$_*$ (K) & 5568$\pm$90& 5246$\pm$80 & 4803$\pm$80&5011$\pm$55&4715$\pm$50\\
Mass (M$_{\tt Jup}$) & 1.949$_{-0.078}^{+0.08}$&0.537$\pm$0.017&0.200$\pm$0.019&0.059$\pm$0.007&0.250$\pm$0.023\\
Radius (R$_{\tt Jup}$) &1.072$\pm$0.043 &1.010$\pm$0.029&0.995$\pm$0.052&0.565$_{-0.032}^{+0.072}$&1.057$\pm$0.047\\
T$_{\tt eq}$ (K)$^{a}$ &902$\pm$27&791$\pm$17&822$\pm$22&1028$\pm$21&961$\pm$20\\
$e^{b}$ &0.200$_{-0.028}^{+0.026}$&0.3417$\pm$0.0036&$<$0.087($<$0.16)&0.14$_{-0.08}^{+0.12}$&$<$0.11($<$0.23)\\
$\omega$ (deg)$^{b}$ &262.5$_{-2.9}^{+2.4}$ &200.5$\pm$1.3 &...&46$^{+33}_{-71}$&...\\
Period (days)$^{c}$&10.863502(37) &10.3385230(90)&5.5080291(42)&4.2345023(15)&3.8681382(17)\\
T$_{c}$(BJD-2,450,000)$^{c}$&4638.56094(48)&4801.17018(20)&4715.02254(39)&5304.65218(25)&5748.83422(18)\\
References  &1,2,3&3,4,5,6 &3,7,8,9&6,10,11,12&3,13

\enddata 
\label{table:systems1}
\tablenotetext{}{\textbf{Notes.}}

\tablenotetext{a}{Calculated assuming planet-wide heat circulation and zero albedo. Uncertainties on the temperature calculated from the uncertainties on a/R$_{*}$ and T$_{*}$. The semi-major axis and R$_{*}$ values and their corresponding uncertainties for HAT-P-15b, HAT-P-17b, and WASP-69b are taken from reference 3 and for HAT-P-18b from reference 9. R$_{*}$ and semi-major axis for HAT-P-26b are from \citet{Wakeford2017HeavyAbundance}. }   
\tablenotetext{b}{The orbital eccentricity $e$ and longitude of periapse $\omega$ are derived from fits to radial velocity data.}  
\tablenotetext{c}{Uncertainties on the last two digits are parenthesized.}  
\tablenotetext{}{\textbf{References.} (1) \citet{Kovacs2010HAT-P-15b:Star}, (2) \citet{Torres2012ImprovedHosts}, (3) \citet{Bonomo2017ThePlanets}, (4) \citet{Howard2012HAT-P-17bc:Jupiter}, (5) \citet{Fulton2013TheSystem}, (6) \citet{Mortier2013NewHosts}, (7) \citet{Hartman2011HAT-P-18bStars}, (8) \citet{Kirk2017RayleighHAT-P-18b}, (9) \citet{Seeliger2015Ground-basedSystems},
(10) \citet{Hartman2011HAT-P-26b:Star}, (11) \citet{Knutson2014FriendsPlanets}, (12) \citet{Stevenson2016ALDSS-3C}, (13) \citet{Anderson2014ThreeBinary}}
\end{deluxetable*}
\end{centering}

In principle we can directly determine the mean molecular weights and corresponding metallicities of transiting planet atmospheres by measuring their wavelength-dependent transit depths or transmission spectra (e.g., \citealt{Seager2000TheoreticalTransits}). However, a majority of the gas giant planets observed to date have clouds in their day-night terminator region that attenuate the amplitude of the expected absorption features (e.g., \citealt{Sing2016ADepletion}; \citealt{Barstow2017ATransmission}), leading to degeneracies between cloud-top pressure and atmospheric metallicity for observations with low signal-to-noise detections (e.g., \citealt{Benneke2012AtmosphericSpectroscopy}).  This problem is especially acute for exo-Neptunes, which typically have smaller planet-star radius ratios and higher surface gravities than their Jovian counterparts, both of which make it more challenging to detect atmospheric absorption during the transit.  There are currently only three exo-Neptunes with published transmission spectra (GJ 436b, \citealt{Knutson2014A436b}; HAT-P-11b, \citealt{Fraine2014WaterExoplanet}; and HAT-P-26b, \citealt{Wakeford2017HeavyAbundance}), and of these three HAT-P-26b is the only one with a relatively clear atmosphere and correspondingly strong constraints on its atmospheric metallicity.  Interestingly, this planet appears to have an atmospheric metallicity substantially lower than that of Neptune \citep{Wakeford2017HeavyAbundance}.

Although clouds are problematic for transmission spectroscopy, observations of the thermal emission spectra of these same cloudy planets indicate the presence of strong molecular absorption features (e.g., HD 189733b, \citealt{Crouzet2014WaterEclipse}, \citealt{Todorov2014Updated189733b}; GJ 436b, \citealt{Morley2017ForwardClouds}).  This is due in part to the shorter path length for thermal emission as compared to transmission spectroscopy, which minimizes the scattering opacity \citep{Fortney2005TheSpectroscopy}.  We also expect that these tidally locked planets should exhibit day-night temperature gradients that might prevent clouds condensing in the cooler terminator region from extending into the hotter dayside region (e.g., \citealt{Demory2013InferenceAtmosphere}, \citealt{Parmentier2016TransitionsJupiters}),  although meridional advection of cloud particles may also affect the observed cloud properties (\citealt{Lee2016}, \citealt{Lines2018}). Secondary eclipse observations of the Neptune-mass planet GJ 436b ($<$800 K) indicate that it has strong molecular features in its emission spectrum that can only be matched by a substantially metal-enriched atmosphere ($200-1000\times$ solar; \citealt{Stevenson2010Possible436b}, \citealt{Moses2013Compositional436b}, \citealt{Lanotte2014AB}, \citealt{Morley2017ForwardClouds}).  

In \citet{Kammer2015SPITZERSPECTRA} we used broadband emission spectroscopy in the same 3.6 and 4.5 $\mu$m bands to constrain the atmospheric compositions of five transiting gas giant planets with temperatures cooler than 1200 K and masses ranging between 0.3 and 3 M$_{\tt Jup}$.  For these relatively cool hydrogen-rich atmospheres, models predict that the ratio of methane to carbon monoxide and carbon dioxide should act as a sensitive barometer of atmospheric metallicity and the carbon-to-oxygen ratio \citep{Moses2013Compositional436b}. \citet{Kammer2015SPITZERSPECTRA} leveraged the fact that the 3.6 $\mu$m \emph{Spitzer} band is sensitive to CH\textsubscript{4} absorption while the 4.5 $\mu$m band is sensitive to CO and CO$_2$ absorption.  The ratio of the measured eclipse depths in these two bands can therefore be used to provide constraints on relative trends in atmospheric composition. Although these measurements hinted at a possible trend in atmospheric metallicity versus planet mass, our sensitivity was limited by the large measurement errors characteristic of these types of observations and our relatively small sample size.
\begin{centering}
\begin{deluxetable*}{ccccccccccc}[t!]
\tabletypesize{\scriptsize}
\tablecaption{\emph{Spitzer} Observation Details}
\tablewidth{0pt}
\tablehead{
\colhead{Target} & \colhead{$\lambda$ ($\mu$m)} & \colhead{UT Start Date} & \colhead{Length (h)} & \colhead{t\textsubscript{int}(s)$^{1}$} &
\colhead{t\textsubscript{trim}(h)$^{2}$} & \colhead{r\textsubscript{pos}$^{3}$ } & \colhead{r\textsubscript{phot}$^{4}$}  & \colhead{n$\textsubscript{bin}$$^{5}$} & RMS$^{6}$}
\startdata
HAT-P-15b&  3.6 &2011 Nov 27 & 7.8&6.0&1.5 & 2.5 & 2.1 & 4& 1.40\\ 
&   3.6 &2012 May 8  &7.9  &2.0 &0.5 &2.5 &2.2 &16  & 1.13\\
&   3.6 &2014 Apr 25 &11.35&2.0 &1.0 &4.0 &2.0 &8  & 1.15\\
&   4.5 &2012 Apr 27 & 7.9 &2.0 &0.0 &4.0 &2.6 &2 & 1.14\\
&   4.5 &2012 Nov 19   &7.9  &2.0 &2.0 &2.5 &2.3 &2   & 1.09\\
&   4.5 &2014 May 27   &11.35 &2.0 &1.5 &4.0 &2.2 &8 & 1.10\\

HAT-P-17b&3.6  &2012 Jan 25 & 7.9 & 2.0  & 0.0&4.0 & 2.0& 2& 1.14\\
&   3.6 &2012 Aug 29 & 7.9 &2.0 & 2.0 & 3.0  & 2.5   & 64 &1.10 \\
&   3.6 &2014 Sep 02 & 8.6 & 2.0& 2.0 & 3.5  &2.4 &4  & 1.21\\
&   4.5 &2012 Feb 04 & 7.9 & 2.0 & 0.5 & 4.0 &2.7 &32 & 1.12\\
&   4.5 &2012 Sep 08 & 7.9 & 2.0 & 2.0 & 3.0 &2.4 &128& 1.10\\
&   4.5 &2012 Sep 22 & 8.6 & 2.0 & 2.0 & 3.5 &2.4 &4  & 1.22\\

HAT-P-18b&   3.6&  2012 May 19 & 7.9& 2.0 &1.5 &4.0 &2.0& 2&1.13 \\
&   3.6 &2014 May 22 &5.8  &2.0 &2.0 &3.0 &2.0 &4  & 1.12\\
&   4.5 &2011 Aug 28 &11.9 &2.0 &1.5 &4.0 &2.5 &64 & 1.15\\
&   4.5 &2014 Jun 2  &5.8  &2.0 &0.5 &4.0 &2.0 &4& 1.09\\

HAT-P-26b& 3.6  & 2014 Apr 11 &7.4  & 2.0 & 1.5 & 2.5 & 2.0& 2& 1.13\\
&   3.6 &2014 Apr 24 &7.4 &2.0 &1.0 &3.5 &2.0&   2& 1.16\\
&   3.6 &2014 Sep 10 &7.4 &2.0 &1.0 &4.0 &2.3& 512& 1.22\\
&   3.6 &2014 Sep 27 &7.4 &2.0 &1.0 &3.5 &2.2&   2& 1.21\\
&   4.5 &2014 Apr 15 &7.4 &2.0 &2.0 &4.0 &2.8&   8& 1.14\\
&   4.5 &2014 May 06 &7.4 &2.0 &1.0 &3.5 &2.4&   2& 1.11\\
&   4.5 &2014 Sep 15 &7.4 &2.0 &0.0 &3.5 &2.0&   2& 1.20\\
&   4.5 &2014 Oct 02 &7.4 &2.0 &2.0 &4.0 &2.1&  64& 1.19\\

WASP-69b &  3.6 & 2014 Jul 22 & 9.8& 0.4 & 2.4 & 3.5 & 2.0& 2& 1.15\\
&   3.6 &2014 Jul 29 &9.8 &0.4 &2.0 &3.5 &2.0 &8 & 1.13\\
&   4.5 &2014 Aug 18 &9.8 &0.4 &0.0 &3.5 &2.3 &2 & 1.16\\
&   4.5 &2015 Jan 08 &9.8 &0.4 &2.0 &3.5 &2.3 &2 & 1.13
\enddata
\tablenotetext{1}{Integration time}
\tablenotetext{2}{Initial trim duration }
\tablenotetext{3}{Radius of the aperture (in pixels) used to determine the location of the star on the array}
\tablenotetext{4}{Radius of the aperture (in pixels) used for the photometry}
\tablenotetext{5}{Bin size used for fits}
\tablenotetext{6}{Ratio of measured RMS to the photon noise limit}
\label{table:observations}
\end{deluxetable*}
\end{centering}

In this study we utilize the Infra-Red Array Camera (IRAC) on board the \emph{Spitzer Space Telescope} to obtain a combined total of 28 3.6 $\mu$m and 4.5 $\mu$m secondary eclipse observations for a sample of five additional transiting gas giant planets with temperatures below $\sim$1000 K and masses between 0.05 and 2.0 M$_{\tt Jup}$ (see Table~\ref{table:systems1} for more information).  Our targets in this study include HAT-P-15b \citep{Kovacs2010HAT-P-15b:Star}, HAT-P-17b \citep{Howard2012HAT-P-17bc:Jupiter}, HAT-P-18b \citep{Hartman2011HAT-P-18bStars}, HAT-P-26b \citep{Hartman2011HAT-P-26b:Star}, and WASP-69b \citep{Anderson2014ThreeBinary}.  Of these five planets, HAT-P-26b is the only one with published constraints on its atmospheric metallicity from transmission spectroscopy \citep{Wakeford2017HeavyAbundance}, with a range of $0.8-26\times$ solar ($1\sigma$). Optical transmission spectroscopy of HAT-P-18b between 475 and 925 nm from the William Herschel Telescope indicates that it has a featureless spectrum consistent with Rayleigh scattering in this wavelength range \citep{Kirk2017RayleighHAT-P-18b}, but this result is still consistent with a wide range of atmospheric metallicities. \citet{Casasayas-Barris2017DetectionWASP-69b} detected sodium absorption at high spectral resolution in the transmission spectrum of WASP-69b, but did not place any constraints on its atmospheric metallicity.

In Section \ref{sec:observations} we describe our photometric extraction and model fits. In Sections \ref{sec:results} and \ref{sec:discussion}, we compare our results to atmosphere models and discuss the corresponding implications for our understanding of trends in atmospheric composition.

\section{Observations and Data Analysis}\label{sec:observations}
\subsection{Photometry and Initial Model Fits}

We obtained a minimum of two visits each in the IRAC 3.6 and 4.5 $\mu$m bands \citep{Fazio2004TheTelescope} for all planets in our sample, with additional observations for lower signal-to-noise targets. A majority of these observations were observed in the $32\times32$ pixel subarray mode with an initial thirty-minute observation to allow for settling of the telescope followed by a peak-up pointing adjustment prior to the start of the science observation \citep{Ingalls2012Intra-pixelIRAC}.  The only exceptions are the 3.6 $\mu$m 2011 November observation of HAT-P-15b, the 3.6 $\mu$m 2012 January and 4.5 $\mu$m 2012 February observations of HAT-P-17b, and the 4.5 $\mu$m 2012 August observation of HAT-P-18b, which did not include this initial 30-minute observation and subsequent pointing adjustment.  The 2011 observation of HAT-P-15b was also obtained in full-array mode instead of subarray mode. See Table~\ref{table:observations} for additional details.
\begin{centering}
\tabletypesize{\scriptsize}
\setlength{\tabcolsep}{3pt}
\begin{deluxetable*}{ccccccc}[t!]
\tablecaption{Best-fit Eclipse Parameters}
\tablewidth{0pt}
\tablehead{
\colhead{Target} & \colhead{Band ($\mu$m)} & \colhead{Depth (ppm)} &\colhead{Brightness Temperature (K)} & \colhead{ Time Offset (days)\textsuperscript{a}} & Center of Eclipse (Phase) & \colhead{ecos($\omega$)$^{b}$} }
\startdata
HAT-P-15b    &  3.6 & $<$ 180 $^c$ & $<$ 971 &0 $\pm$0.0585 $^{f}$ &0.4829$\pm$0.0054 $^{f}$& -0.0262$^{+0.0082}_{ -0.0084}$ $^{g}$\\
              &  4.5 & $<$931 $^{c, d}$ & $<$1355  & \\
 HAT-P-17b    &  3.6 & 118 $\pm$ 39&813$_{-61}^{+49}$  &0.0120 $_{-0.0130}^{ +0.0120}$ &0.2997$\pm$0.0012 & -0.3146$\pm$0.0018\\
 &  4.5 & $<$ 149 $^c$ & $<$ 708  & \\
 HAT-P-18b  &  3.6 & 437 $_{-144}^{ +146}$ $^e$& 1004$_{-94}^{+78}$ & 0.0091  $_{-0.0073}^{ +0.0054}$ &0.5016$_{-0.0013}^{ +0.0010}$ & 0.0026$_{-0.0021}^{ +0.0015}$\\
&   4.5 & 326 $_{-147}^{ +144}$ $^e$ &783$_{-100}^{+77}$  \\
HAT-P-26b  &  3.6 & $<$ 85 $^c$ & $<$949 &0.0050$\pm$0.0037  & 0.5012 $\pm$ 0.0009& 0.0019$\pm$0.0014\\
&  4.5 & 265$_{-72}^{+68}$   &1087  $_{-102}^{ +91}$&0.0045$_{-0.0038 }^{+0.0031}$ &0.5011$_{-0.0009}^{ +0.0007}$ & 0.0017$\pm$0.0011 \\
WASP-69b  &  3.6 & 421$\pm$29& 1011$\pm$17 &0.0033 $\pm$ 0.001& 0.5009$\pm$0.0003& 0.001$\pm$ 0.0004\\
 &   4.5 & 463$\pm$39   & 863$_{-20}^{+19}$
\enddata 
\tablenotetext{a}{Time offset from the predicted center of the eclipse.  Unless otherwise noted, we fit both channels with a common time of secondary eclipse.}
{\tablenotetext{b}{Computed using the approximation for a low eccentricity orbit. Therefore, this only serves as a first-order approximation for the more highly eccentric orbit of HAT-P-17b. See \cite{Pal2010} for a detailed discussion of the correct treatment for higher eccentricity orbits.}}
\tablenotetext{c}{We report the 2$\sigma$ upper limit for the eclipse depth.}
{\tablenotetext{d}{The solution for the eclipse depth at 4.5$\mu$m was multimodal, so we report the 2$\sigma$ upper limit corresponding to the deepest of the eclipse solutions to be conservative. See Section \ref{sec:fits} for more details.}}
\tablenotetext{e}{We report the secondary eclipse depths using a tight uniform prior (see Section \ref{sec:fits} for more details).}
{\tablenotetext{f}{Due to the nondetections in both bandpasses, we simply report the time offset and corresponding phase of the Gaussian prior derived from the RV constraints from Bonomo et al. (\citeyear{Bonomo2017ThePlanets}; see Section \ref{sec:fits} for more details). }}
{\tablenotetext{g}{We report the ecos($\omega$) from \cite{Bonomo2017ThePlanets}. }}
\label{table:bestfit}
\end{deluxetable*}
\end{centering}

We utilize the standard Basic Calibrated Data (BCD) images for our analysis and extract photometric fluxes as described in our previous studies (i.e., \citealt{Knutson2008TheInversion}; \citealt{Kammer2015SPITZERSPECTRA}; \citealt{Morley2017ForwardClouds}).  We first calculate the BJD\textsubscript{UTC} mid-exposure times for each image.  We then estimate the sky background in each image by masking out a circular region with a radius of 15 pixels centered on the position of the star, iteratively trimming $3\sigma$ outliers, and fitting a Gaussian function to a histogram of the remainder of the pixels. For the full-array observation of HAT-P-15b, we determine the sky background using the median flux in an annulus with radii between 15 and 37 pixels centered on the position of the star.  We determine the location of the star on the array using flux-weighted centroiding (e.g., \citealt{Knutson2008TheInversion}; \citealt{Deming2015SpitzerDecorrelation}) with a circular aperture.  We consider aperture radii ranging between 2.5 and 4.0 pixels in 0.5 pixel steps and optimize our choice of aperture as described below.  We then extract the total flux in a circular aperture centered on the position of the star using the \texttt{aper} routine in the \texttt{DAOPhot} package \citep{Stetson1987DAOPHOT:APhotometry}. We consider aperture sizes ranging from 2.0 to 3.0 pixels in steps of 0.1 pixels and from 3.0 to 5.0 pixels in steps of 0.5 pixels.  

Some visits also display a ramp-like behavior at early times, which we mitigate by trimming up to two hours from the start of our time series.  As discussed in \citet{Deming2015SpitzerDecorrelation} and \citet{Kammer2015SPITZERSPECTRA}, we find that binning our data before fitting reduces the amount of time-correlated noise in the residuals.  We determine the optimal flux-weighted centroiding aperture, photometric aperture, trim duration, and bin size for each visit by first fitting a combined instrumental and astrophysical model to each version of the photometry and then calculating the standard deviation of the residuals as a function of bin size stepping in powers of two as described in \citet{Kammer2015SPITZERSPECTRA}. We then calculate the least-squares difference between the measured standard deviation of the residuals and the predicted photon noise limit in each bin, which decreases as the square root of the number of points.  We then select the photometric and centroiding apertures, trim duration, and bin size that minimizes this least-squares difference (i.e., the one that is closest to the photon noise at all measured timescales) for use in our subsequent analysis.  Because we typically do not detect the eclipse in each individual visit, we fix the time of secondary eclipse to the predicted value (phase=0.5 for HAT-P-18b, HAT-P-26b, and WASP-69b, and using the best-fit radial velocity (RV) solution for HAT-P-15b and HAT-P-17b) during our initial optimization (see Section \ref{sec:fits} for additional details). Our observations of WASP-69b also showed a steep downward trend after the end of the eclipse, which was not well matched with our standard linear function of time. In order to avoid fitting a quadratic function of time, which has the potential to bias our measured eclipse depth, we also trimmed up to 2 hours from the end of the time series for all of the WASP-69b observations.  We optimized this trim duration in the same manner as for the initial trim duration.

Our model for each visit consists of an eclipse model and an instrumental noise model, which we fit simultaneously.  We calculate our eclipse model using the \texttt{batman} package (\citealt{Mandel2002AnalyticSearches}, \citealt{Kreidberg2015Batman:Python}), 
where we fix the planet-star radius ratio, orbital inclination, and the ratio of the orbital semi-major axis  to the stellar radius ($a/R_*$) to the published values for each planet (see Table ~\ref{table:systems1} for references) and allow the eclipse depth and time to vary as free parameters.

The flux we measure also depends on the position of the star on the array in each image. This is due to \emph{Spitzer}'s well-documented intrapixel sensitivity variations (e.g., \citealt{Charbonneau2005DetectionPlanet}; \citealt{Reach2005AbsoluteTelescope}; \citealt{Morales-Calderon2006AWeather}) combined with an undersampled stellar point spread function (the FWHM in the 3.6 and 4.5 $\mu$m arrays is approximately two pixels for data taken during the post-cryogenic mission).  We correct for this effect using the pixel-level decorrelation (PLD) method \citep{Deming2015SpitzerDecorrelation}. This method uses a linear combination of individual pixel-level light curves as the instrumental noise model, and is therefore able to capture trends due to variations in both the star's position and the width of the stellar point spread function. As in \citet{Deming2015SpitzerDecorrelation}, we utilize a 3 x 3 grid of pixels centered on the location of the star in our model. We remove astrophysical flux variations in each 3 x 3 postage stamp by dividing by the summed flux across all nine pixels.  Our final instrumental noise model therefore consists of nine linear coefficients corresponding to the nine individual pixel-level light curves, as well as a linear function of time to capture any long-term trends (i.e., 11 free parameters in total).  For all fits we divide out our initial astrophysical model and use linear regression on the residuals to obtain an initial guess for the nine linear PLD coefficients in order to speed up convergence for these highly correlated parameters.

\begin{centering}
\begin{figure*}[t!]
\includegraphics[width=\textwidth]{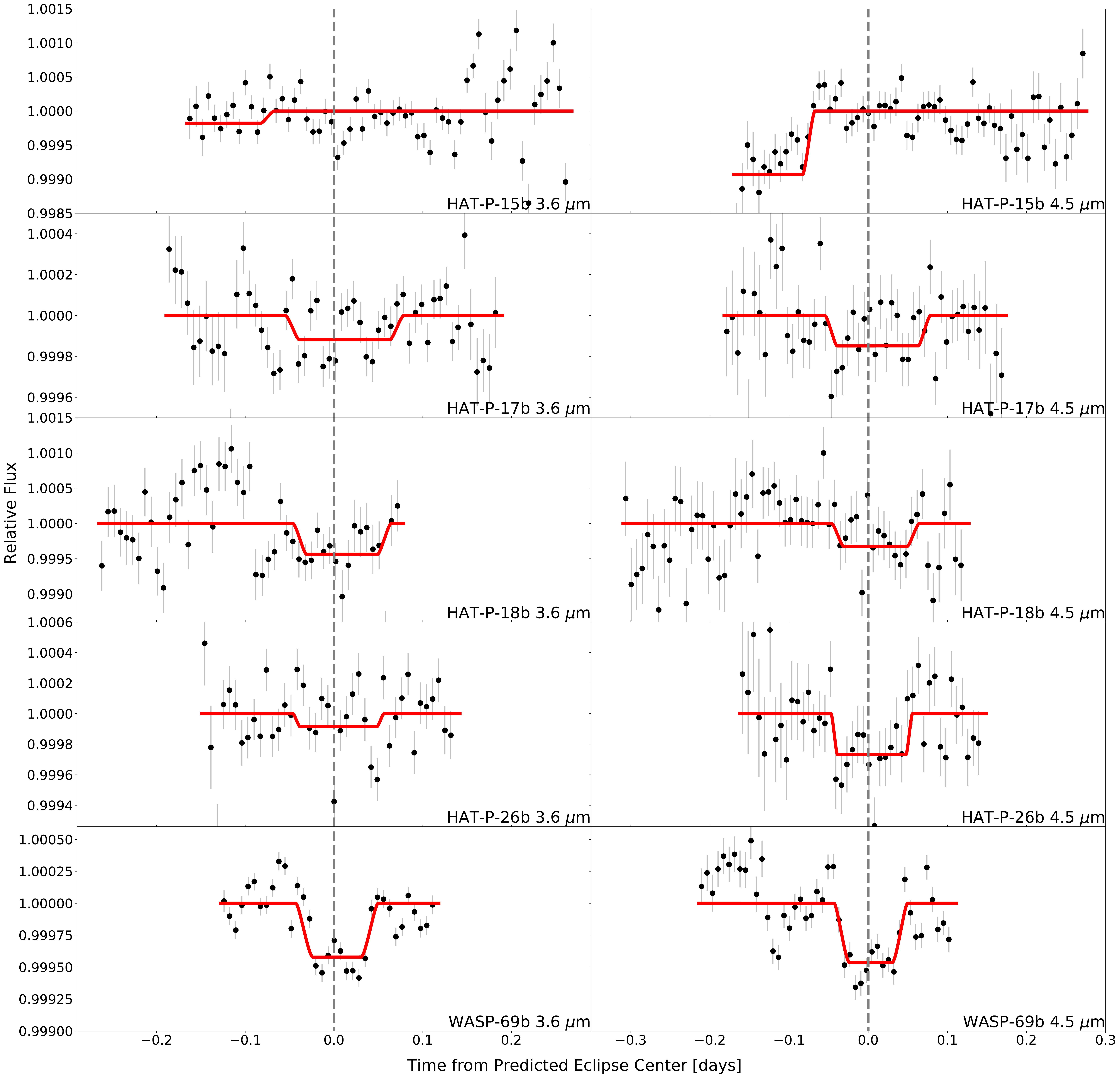}
\caption{Phased light curves for each planet from the simultaneous fits with instrumental effects removed. We show data binned in ten-minute intervals (black filled circles) with error bars corresponding to the scatter in each bin, and overplot the best-fit eclipse model in each bandpass for comparison (red lines). The 2$\sigma$ upper limits for the best-fit eclipse depths of the 3.6 $\mu$m and 4.5 $\mu$m bandpasses for HAT-P-15b (see Section \ref{sec:fits} for more details), the 4.5 $\mu$m bandpass for HAT-P-17b, and the 3.6 $\mu$m bandpass for HAT-P-26b are shown. }
\label{fig:light_curves}
\end{figure*}
\end{centering}

\subsection{Simultaneous Fits and Choice of Prior on Eclipse Phase}\label{sec:fits}

Because a majority of the planets in our study have relatively shallow eclipse depths, we do not expect to detect the eclipse signal in fits to individual visits.  We therefore carry out our initial fits to individual visits using a fixed eclipse time.  For HAT-P-18b, HAT-P-26b, and WASP-69b the published RVs are consistent with a circular orbit (\citealt{Knutson2014FriendsPlanets}; \citealt{Bonomo2017ThePlanets}) and we therefore fix the eclipse phase to 0.5. HAT-P-15b and HAT-P-17b have non-zero orbital eccentricities and we therefore fix the eclipse time to the predicted value from the literature \citep{Bonomo2017ThePlanets}.  After optimizing our choice of aperture, bin size, and trim duration for each individual visit, we next carry out a joint fit to all of the visits for a given planet.  In these fits we assume a common eclipse depth for each bandpass and a common eclipse phase for all visits regardless of bandpass, and allow these three parameters to vary in our fits.  In this case we place a uniform prior on the eclipse time spanning the range of times where the full eclipse would be visible in the data (i.e., we disallow eclipse times that are either partially or fully outside our observational window).  We then fix the eclipse time to the best-fit value from the simultaneous fit and revisit our choice of optimal aperture, bin size, and trim duration for each individual visit. Lastly, we rerun the simultaneous fit using these newly optimized light curves.

We estimate the uncertainties on the best-fit eclipse parameters in these simultaneous fits using the affine-invariant Markov chain Monte Carlo (MCMC) ensemble sampler \texttt{emcee} \citep{Foreman-Mackey2013Emcee:Hammer}.  Our combined astrophysical and instrumental noise model has 13 free parameters, and we therefore use 60 walkers in our fits in order to ensure sufficient sampling of the model parameter space.  We place uniform priors on all of our model parameters except where noted below.  We also allow the eclipse depths to take on negative values in our fits (i.e., an increase in flux during the eclipse) in order to avoid biasing our estimate of the eclipse depth by requiring only positive values.  We initialize the walkers in a tight cluster centered on the best-fit solution from a Levenberg-Marquardt minimization and carry out an initial burn-in with a length of 10,000 steps.  We then discard this initial burn-in and carry out a subsequent fit with 10$^5$ steps per chain. We report the median values from our MCMC chains and the corresponding 1$\sigma$ uncertainties. 

We show the raw photometry for each visit with best-fit instrumental noise models from the simultaneous fit overplotted in Figures~\ref{fig:A1} -~\ref{fig:A3} in the appendix.  Normalized light curves for these visits with best-fit eclipse light curves overplotted are shown in Figures ~\ref{fig:A4} -~\ref{fig:A6}.  The standard deviation of the residuals as a function of bin size for each visit is shown in Figures ~\ref{fig:A7} -~\ref{fig:A9}, with the predicted photon noise limit for each bin size overplotted for comparison.  We also combine all visits within the same bandpass and show these averaged light curves in Figure ~\ref{fig:light_curves}.

We modified our approach for HAT-P-26b, which is the only planet in our sample with four visits in each bandpass.  In this case a simultaneous fit to all eight visits would require a prohibitively large model with a total of 91 free parameters.  We instead elect to fit each bandpass separately, and find that the eclipse is detected at 4.5 $\mu$m but not at 3.6 $\mu$m.  We therefore repeat our fits to the 3.6 $\mu$m data with a Gaussian prior on the eclipse time centered on the best-fit eclipse phase from the 4.5 $\mu$m fits and with a width equal to the $\pm1\sigma$ uncertainty on this parameter.

HAT-P-15b was the only planet in our sample with no eclipse detection in either band.  Previous RV observations of this planet indicate that it has an orbital eccentricity of $0.190\pm0.019$ \citep{Kovacs2010HAT-P-15b:Star}, and we therefore centered our window on the predicted secondary eclipse phase rather than a phase of 0.5. \cite{Bonomo2017ThePlanets} subsequently reported an updated eccentricity constraint of $0.200^{+0.026}_{-0.028}$ with a corresponding uncertainty in the predicted eclipse phase of $\pm0.058$ days.  This means that our shortest observational window for this planet (7.8 hours) encompassing the entirety of the eclipse only spanned -1.5$\sigma$ to +1.3$\sigma$, while the longest observation (11 hours) spanned -2.0$\sigma$ to +3.4$\sigma$. Because we do not expect to detect the eclipse at a statistically significant level in a single visit, the shortest observation window becomes the limiting factor on the effective phase range of our search.

\begin{centering}
\begin{figure}[h!]
\includegraphics[height=.45\textwidth]{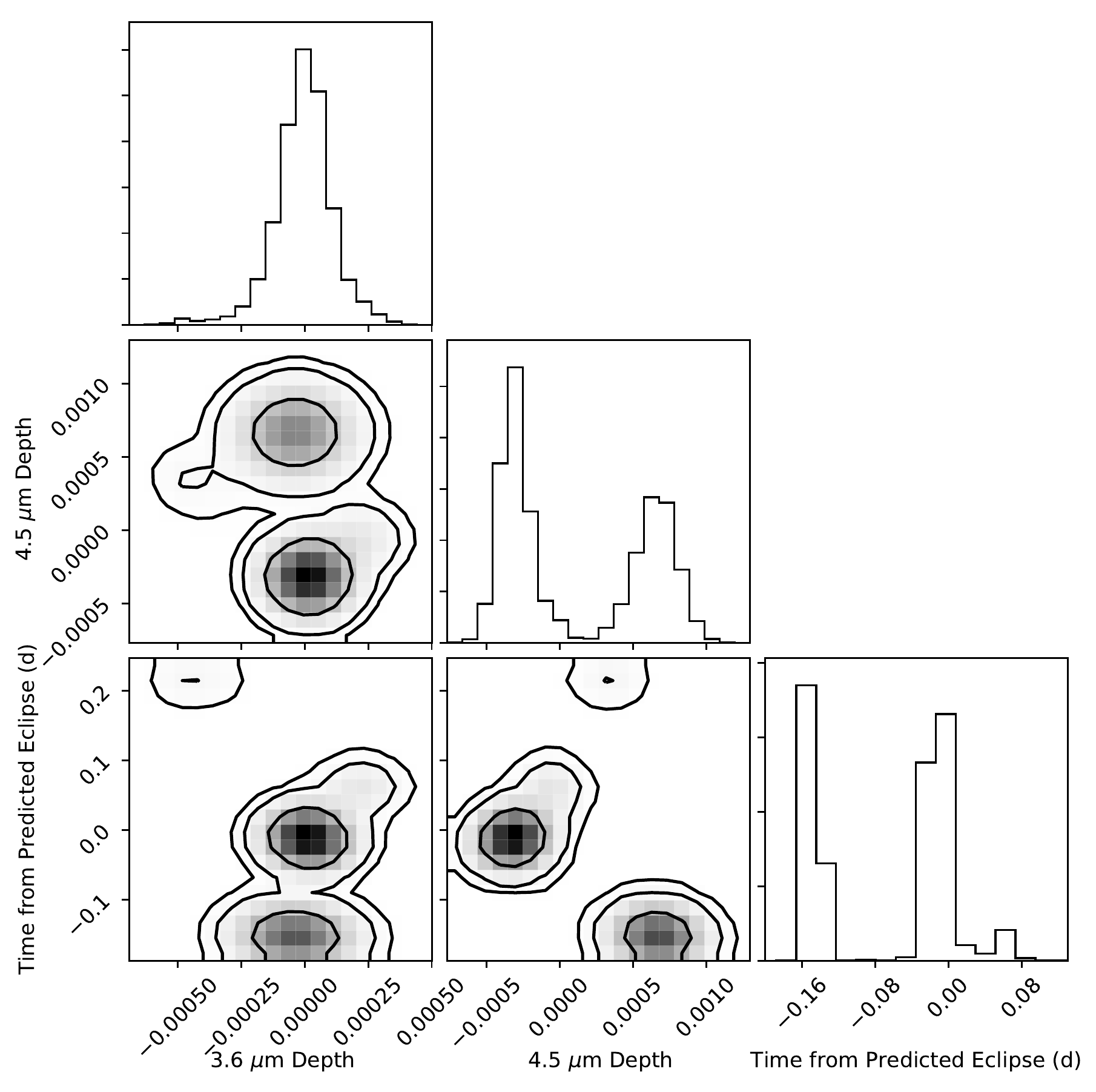}
\caption{Posterior probability distribution for the secondary eclipse center time and depth in both bands from a joint fit with a Gaussian prior derived from the RV constraints on the eclipse center time for HAT-P-15b. Contours indicate the 1$\sigma$, 2$\sigma$, and 3$\sigma$ bounds on these parameters.}
\label{fig:h15}   
\end{figure} 
\end{centering}

Within this range, we place an upper limit on the eclipse depth in each bandpass by carrying out fits with a Gaussian prior on the eclipse phase based on the RV fit from \citet{Bonomo2017ThePlanets}. This results in a multimodal posterior for the best-fit eclipse center time, with one peak corresponding to a fit in which the center of the secondary eclipse occurred at the very beginning of the observations (i.e., the entirety of the eclipse is not within the data) and the other peak centered at the expected time of secondary eclipse (see Figure \ref{fig:h15}). The fitted 3.6 $\mu$m secondary eclipse depth was consistent with zero in both peaks, but the 4.5 $\mu$m secondary eclipse depth was bimodal. In order to be more conservative, we report the the $2\sigma$ upper limits on the eclipse depths corresponding to the peak that is centered at the earlier time of secondary eclipse (i.e. the time that gives a positive upper limit on the 4.5 $\mu$m eclipse depth).

We also test how the use of a Gaussian prior on the time of secondary eclipse as determined by RV measurements effects the measured eclipse depths for HAT-P-17b. We find that using either a flat prior (the time of secondary eclipse must be centered between -0.1 days and 0.1 days where zero is the time of secondary eclipse predicted using the \citet{Bonomo2017ThePlanets} eccentricity constraints) or a Gaussian prior (with the mean of the distribution occurring at the time of the secondary eclipse corresponding to zero days and a standard deviation of 1$\sigma$ as determined from the \citet{Bonomo2017ThePlanets} eccentricity constraints) results in the same measured eclipse depths to within 1$\sigma$. Therefore, we report the best-fit values using the less restrictive prior.

Our observations of HAT-P-18b also proved to be particularly challenging.  While \citet{Hartman2011HAT-P-18bStars} reported a non-zero eccentricity for HAT-P-18b with $\sim$ 2$\sigma$ significance, we subsequently acquired additional RV measurements and refit these data in \citet{Knutson2014FriendsPlanets}, where we found that the orbit was consistent with zero eccentricity (e$=0.11^{+0.15}_{-0.08}$). As a result, we centered the 2011 4.5 $\mu$m and 2012 3.6 $\mu$m observations on the predicted eclipse time for the eccentric orbit from \citet{Hartman2011HAT-P-18bStars}, and then centered the subsequent 2014 3.6 and 4.5 $\mu$m observations on an orbital phase of 0.5 (i.e., a circular orbit). Although the 2011 4.5 $\mu$m observation is not centered on a phase of 0.5, it does contain the entirety of the eclipse detected in our simultaneous fits (see Figure ~\ref{fig:A5}). However, the 2012 3.6 $\mu$m observation only spans the first half of the eclipse.

\begin{centering}
\begin{figure}[h]
\includegraphics[height=.45\textwidth]{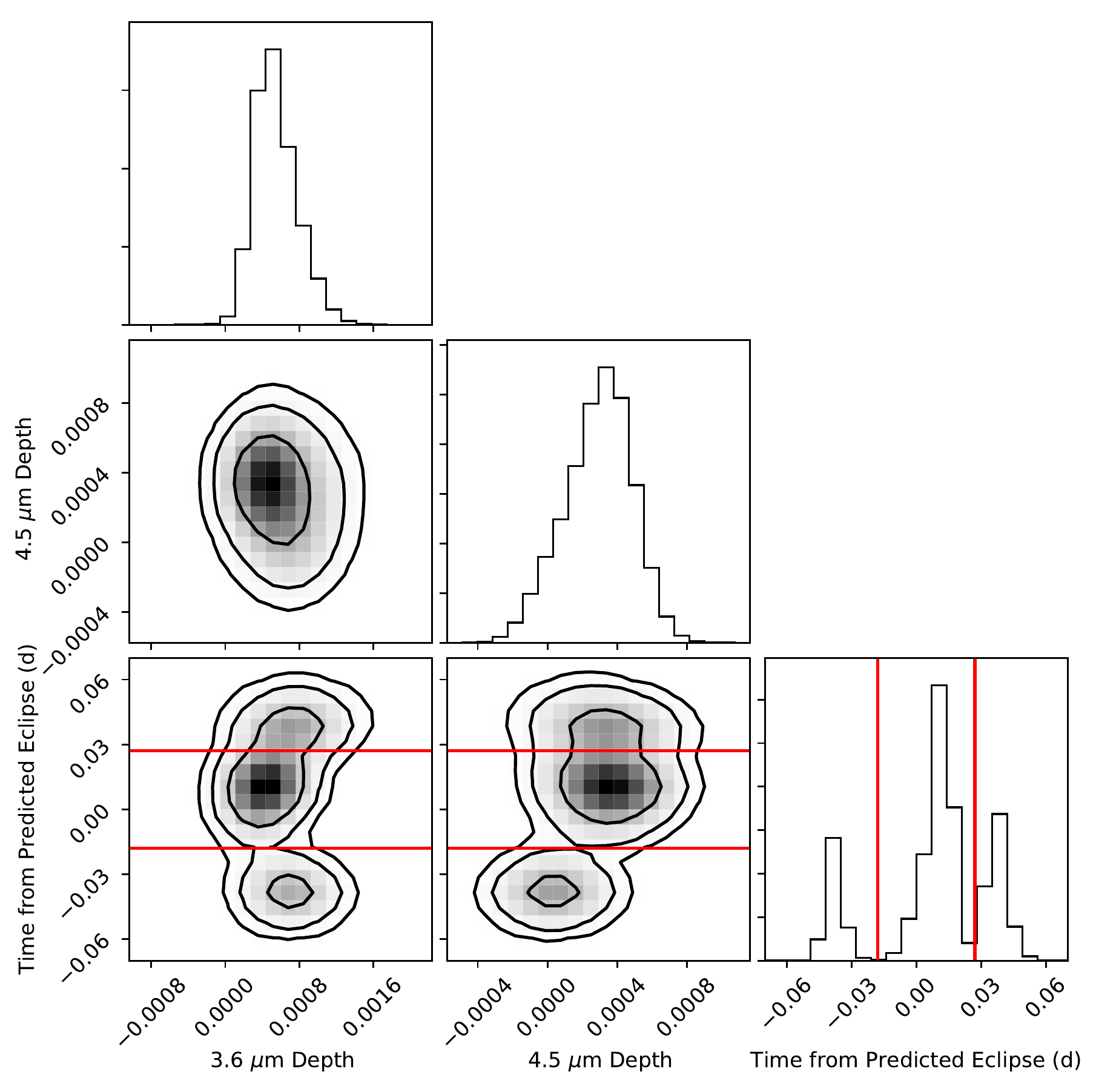}
\caption{Posterior probability distribution for the secondary eclipse center time and depth in both bands from a joint fit with a relatively broad uniform prior of -0.07 to 0.07 days on the eclipse center time for HAT-P-18b. Contours indicate the 1$\sigma$, 2$\sigma$, and 3$\sigma$ bounds on these parameters, while the red lines show the prior constraints used in our final version of the fits for this planet.}
\label{fig:h18}   
\end{figure} 
\end{centering}

When we tried to fit the two channels of HAT-P-18b separately as we did for HAT-P-26b, we found marginal 2.7 and 1.2$\sigma$ detections in the 3.6 $\mu$m and 4.5 $\mu$m bandpasses, respectively, when tight flat priors were used. However, switching to a joint fit of both bands with a flat prior allowing for the secondary eclipse to occur any time during the full range of orbital phases spanned by our observations resulted in a multimodal posterior on the best-fit eclipse center time (see Figure \ref{fig:h18}) with marginal eclipse depths for both channels ($<$1$\sigma$).

The highest peak in this distribution is centered near zero, corresponding to an eclipse center time consistent with an orbital phase of 0.5.  If we instead repeat our fits using a tighter uniform prior of $-0.018$ days to 0.027 days, therefore excluding the other two weaker peaks, it increases the significance of the detection in both bands to 3.0 and 2.2$\sigma$, respectively.  We adopt this version of the fits as our final solution, as there is currently no evidence for a non-zero orbital eccentricity in the RV data for this planet and this peak had the highest posterior probability in our original fit. 

\section{Results}\label{sec:results}

We report the best-fit eclipse depths and times and their corresponding uncertainties in Table~\ref{table:bestfit}.  We detect the eclipse in both channels with high significance for WASP-69b and with somewhat lower significance for HAT-P-18b.  We detect the eclipse at 4.5 $\mu$m but not at 3.6 $\mu$m for HAT-P-26b, detect the eclipse at 3.6 $\mu$m but not at 4.5 $\mu$m for HAT-P-17b, and do not detect the eclipse in either channel for HAT-P-15b.

In order to interpret our results, we first convert the measured eclipse depth in each bandpass to a brightness temperature (e.g. \citealt{Schwartz2015BalancingAbsorbers}). 
We then check for differences in brightness temperature between bands, which are indicative of changes in the shape of the planet's emission spectrum due to molecular features (see Section \ref{sec:Tbright}).  We find that WASP-69b has molecular features (i.e., nonblackbody emission spectra) detected with a significance greater than $3\sigma$, while HAT-P-17b, HAT-P-18b, and HAT-P-26b differ from the blackbody model by less than $3\sigma$. 

These same brightness temperatures can also be used to estimate the efficiency of heat recirculation between the planet's day- and nightsides. We find that the band-averaged brightness temperatures for all four planets with detected eclipses are consistent with their respective equilibrium temperatures (calculated assuming efficient day-night circulation and zero albedos), suggesting that they have either efficient day-night circulation, non-zero albedos, or a combination of the two (\citealt{Kammer2015SPITZERSPECTRA}; \citealt{Schwartz2015BalancingAbsorbers}, \citeyear{Schwartz2017KnotPhotometry}).

We next use our best-fit eclipse phases to place tighter constraints on the values of ecos($\omega$) for each planet. HAT-P-18b, HAT-P-26b, and WASP-69b all have time offsets that are consistent with a circular orbit (within $\sim$3$\sigma$ of the eclipse time predicted from a circular orbit). HAT-P-17b was previously known to be eccentric, and our new observations confirm and refine the published eccentricity and longitude of periastron from \citet{Bonomo2017ThePlanets}. 

\section{Discussion}\label{sec:discussion}
\subsection{Comparison to 1D Atmosphere Models}
We compare our best-fit eclipse depths to predictions from 1D atmosphere models. Briefly, these models calculate the temperature structure of the atmosphere assuming both chemical and radiative-convective equilibrium. These models are described in more detail in  \citeauthor{Fortney2008AAtmospheres} (\citeyear{Fortney2005TheSpectroscopy}, \citeyear{Fortney2008AAtmospheres}) and \citeauthor{Morley2017ForwardClouds} (\citeyear{Morley2013Quantitatively1214b},  \citeyear{Morley2017ForwardClouds}). Cross sections for molecular and atomic species are described in detail in \citeauthor{Freedman08} (\citeyear{Freedman08}, \citeyear{ Freedman14}), with updates to several species that are described in Marley et al.(2019, in preparation). Moderate resolution spectra are calculated using the thermal emission code described in the appendix of \citet{Morley2015}. Stellar spectra are calculated using PHOENIX model atmospheres for the stellar properties given in Table \ref{table:systems1}. Surface gravities are calculated using the planet masses and radii in Table \ref{table:systems1}. To calculate the incident flux on the planet, we assume that the planet-star distance is the semimajor axis for both circular and eccentric planetary orbits. We assume that heat is either efficiently redistributed or inefficiently redistributed to the nightside. We calculate model spectra for a range of metallicities from solar to 100$\times$ solar metallicity and a range of C/O ratios from C/O=0.15 to 1.5. 

We also develop 1D thermo/photochemical kinetics and transport models for these planets to investigate the possible effects of disequilibrium chemistry (i.e., transport-induced quenching and photochemistry) on the atmospheric composition.  These models use the Caltech/JPL KINETICS code \citep{allen81} to solve the continuity equations for 92 neutral H-, C-, O-, and N-bearing species that interact via $\sim$1600 forward-reverse chemical reaction pairs.  The reaction list is derived from \citet{Moses2013Compositional436b}, and further details of the exoplanet disequilibrium-chemistry modeling can be found in Moses et al. (\citeyear{moses11}, \citeyear{ Moses2013Compositional436b}, \citeyear{moses16}).  Vertical transport in the models occurs through molecular and eddy diffusion, with the vertical profile of the eddy diffusion coefficient ($K_{zz}$) assumed to be similar to that derived for HD 189733b from general circulation models \citep{agundez2014pseudo2D}.  Specifically, $K_{zz}$ = 1$\times 10^7/(P (\textrm{bar}))^{0.65}$ cm$^{2}$ s$^{-1}$ in the radiative region in the upper troposphere and middle atmosphere (restricted to never exceeding 10$^{10}$ cm$^{2}$ s$^{-1}$ in the upper atmosphere), and a constant-with-altitude value of 10$^{10}$ cm$^{2}$ s$^{-1}$ in the convective region at pressures $P$ greater than 100 bar.  

The vertical grid in the disequilibrium model consists of 198 levels separated uniformly in log pressure.  The thermal structure is taken from the radiative-convective equilibrium models described above.  Zero flux boundary conditions are assumed at the top and bottom boundaries, and chemical-equilibrium abundances are assumed for the initial conditions.  The protosolar abundances from Table 10 of \citet{lodders09} are assumed to be representative of solar composition, but the models assume that 20.7\% of the oxygen is removed at depth as a result of the formation of silicates and other refractory condensates \citep[see][]{visscher10rock}.  The solar spectrum at solar minimum is adopted for the stellar ultraviolet spectrum for planets with G- and K-type host stars and the composite M-type stellar spectrum described in \citet{ Moses2013Compositional436b} is adopted for planets with M-type host stars.  All fluxes are scaled to the appropriate planet-star distance.

\begin{centering}
\begin{figure*}[h!]
\begin{centering}
\includegraphics[width=.87\textwidth]{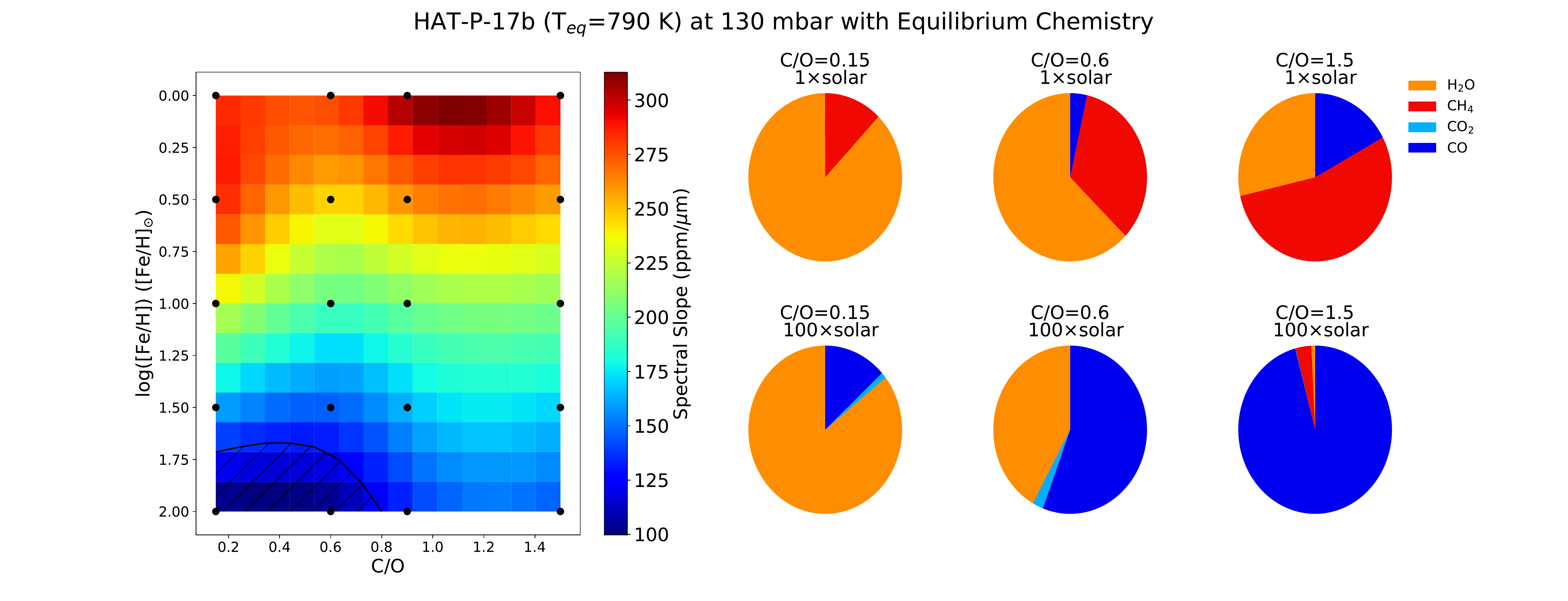}
\caption{Heat map (left) showing the spectral slope expected for equilibrium chemical models with varying C/O ratios and metallicities for HAT-P-17b. Each black point represents a forward model, where we interpolate between models to generate the heat map. We indicate the region of parameter space consistent with this planet's measured spectral slope at the $2\sigma$ level or better with black diagonal lines. The pie charts (right) show the abundances of H$_{2}$O, CH$_{4}$, CO$_{2}$, and CO at a pressure representative of those probed in our observations for select models.}
\label{fig:H17_contour}   
\end{centering}
\end{figure*}
\end{centering}

\begin{centering}
\begin{figure*}[h!]
\begin{centering}
\includegraphics[width=.87\textwidth]{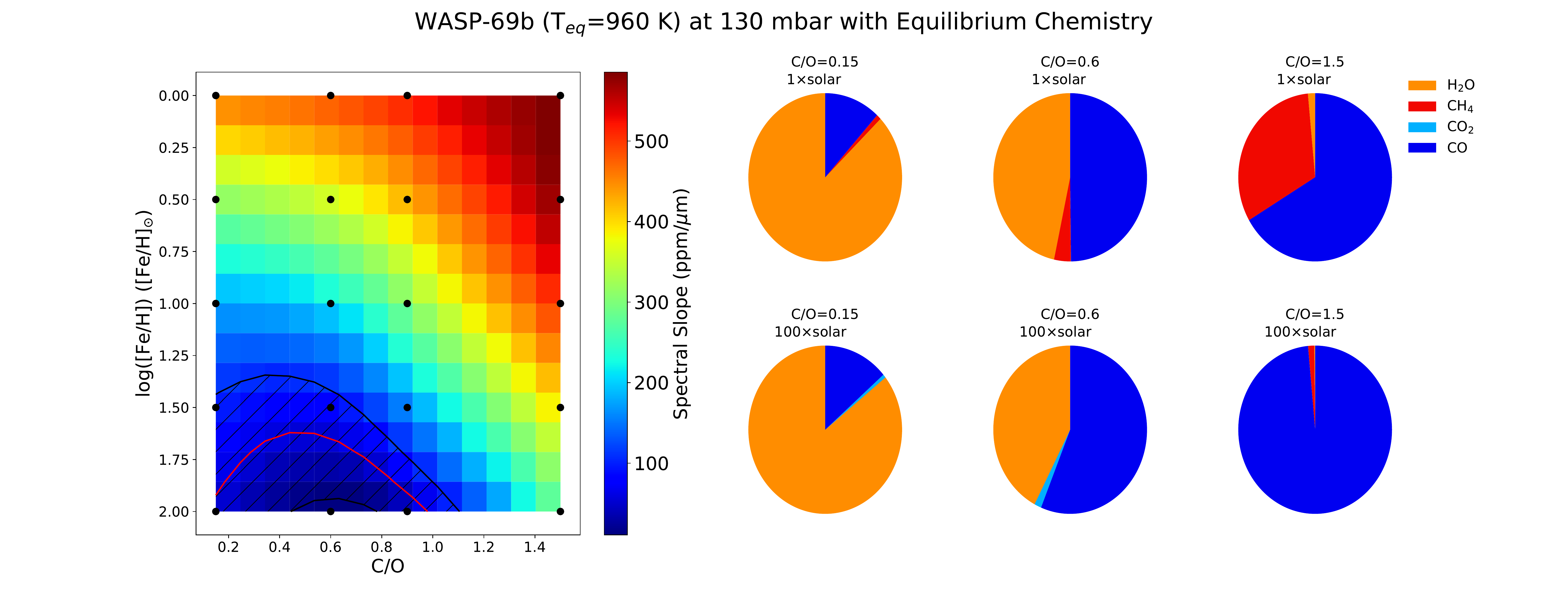}
\caption{Heat map (left) showing the spectral slope expected for equilibrium chemical models with varying C/O ratios and metallicities for WASP-69b. Each black point represents a forward model, where we interpolate between models to generate the heat map. We show the best-fit value of this planet's measured spectral slope in red and show the region of parameter space consistent at the $1\sigma$ level or better with black diagonal lines. The pie charts (right) show the abundances of H$_{2}$O, CH$_{4}$, CO$_{2}$, and CO at a pressure representative of those probed in our observations for select models.}
\label{fig:W69_contour}   
\end{centering}
\end{figure*}
\end{centering}

\begin{centering}
\begin{figure*}[h!]
\begin{centering}
\includegraphics[width=.87\textwidth]{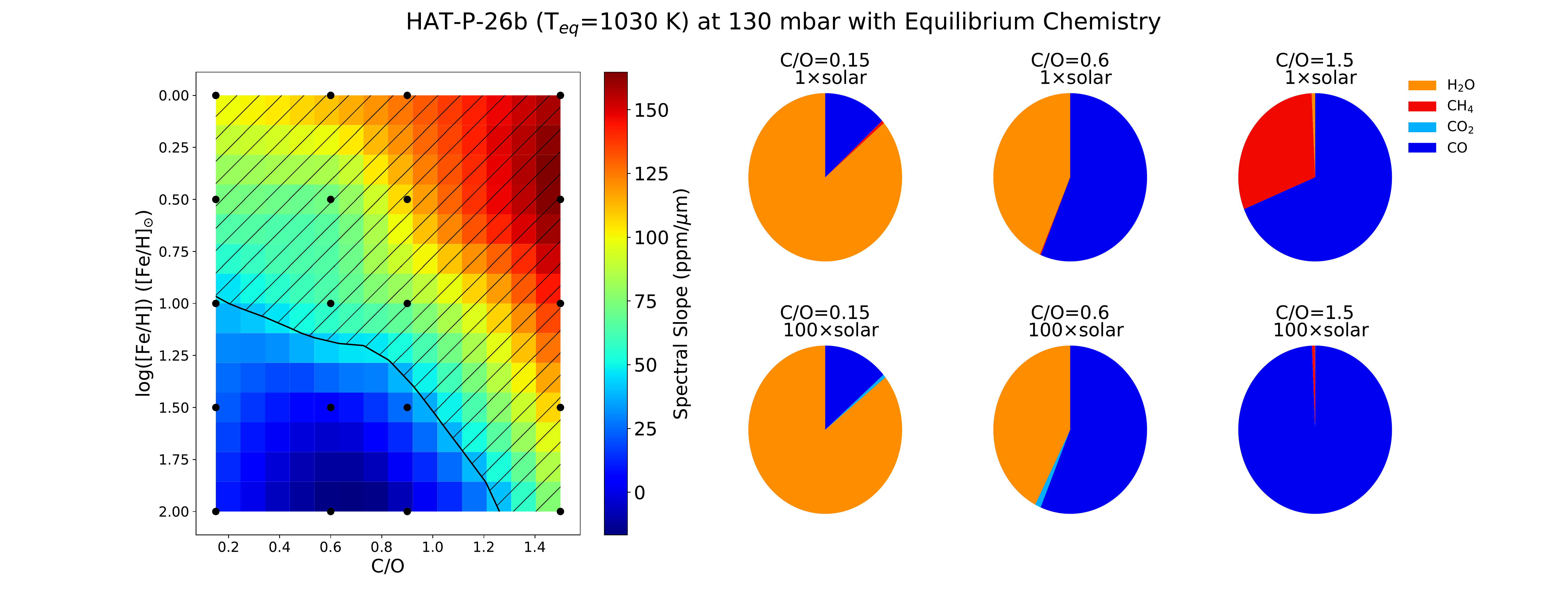}
\caption{Heat map (left) showing the spectral slope expected for equilibrium chemical models with varying C/O ratios and metallicities for HAT-P-26b. Each black point represents a forward model, where we interpolate between models to generate the heat map. We indicate the region of parameter space consistent with this planet's measured spectral slope at the $2\sigma$ level or better with black diagonal lines. The pie charts (right) show the relative abundances of H$_{2}$O, CH$_{4}$, CO$_{2}$, and CO at a pressure representative of those probed in our observations for select models.}
\label{fig:H26_contour}   
\end{centering}
\end{figure*}
\end{centering}

\subsubsection{Chemistry of Cool Hydrogen-rich Atmospheres}\label{sec:chemistry}
We first use these models to examine the effect that varying atmospheric metallicity and the carbon-to-oxygen ratio has on the measured 3.6 and 4.5 $\mu$m broadband fluxes from these planets, in order to determine the degree of degeneracy between these two parameters.  This topic has been previously explored with an extensive grid of generic equilibrium chemistry models in \cite{Molaverdikhani2018}, and also with both forward modeling and atmospheric retrievals for the specific case of GJ 436b \citep{Moses2013Compositional436b,Morley2017ForwardClouds}.  These studies indicate that GJ 436b's very low ratio of CH$_{4}$ to CO and CO$_2$ can only be reproduced by relatively high ($>$200$\times$ solar) metallicity atmospheres even when the C/O ratio is allowed to vary as a free parameter.  In this study we focus on results using forward models, as our two broadband data points are not sufficient for a full retrieval. 

We run a grid of equilibrium models with C/O ratios of [0.15, 0.6, 0.9, 1.5] and metallicities of [1, 3, 10, 30, 100] for the hottest (HAT-P-26b; T$_\texttt{eq}=1030\pm20$ K), the coolest (HAT-P-17b; T$_\texttt{eq}=790\pm20$ K, comparable to GJ 436b), and intermediate temperature (WASP-69b; T$_\texttt{eq}=960\pm20$ K) planets in our sample and determine the expected 3.6 to 4.5 $\mu$m spectral slopes from each of these models. As shown in Figures  \ref{fig:H17_contour}-\ref{fig:H26_contour} these models indicate that the smallest 3.6 to 4.5 $\mu$m spectral slopes (corresponding to planets that are relatively bright at 3.6 $\mu$m and dim at 4.5 $\mu$m) can only be achieved by models with both relatively high ($>50\times$ solar) metallicities and C/O ratios less than $\sim1.5$. This is because at solar C/O the mixing ratios of H$_{2}$O, CH$_{4}$, CO, and CO$_{2}$ all increase as the atmospheric metallicity increases from 1$\times$ to 100$\times$ solar, leading to greater atmospheric opacity, higher temperatures, and a higher overall emission flux in the continuum regions. However, the CO and CO$_{2}$ mixing ratios increase much more rapidly with increasing metallicity than the CH$_{4}$ mixing ratio, leading to a significantly lower CH$_4$/(CO + CO$_{2}$) ratio and greater absorption in the 4.5 $\mu$m band (where CO and CO$_{2}$ absorb) than in the 3.6 $\mu$m band (where CH$_{4}$ absorbs). 

This overall picture remains true for all but the most extreme C/O ratios.  Decreasing the atmospheric C/O ratio at 1$\times$ solar metallicity from 0.9 to 0.15 on a relatively warm planet like HAT-P-26b has only a small effect on the dominant oxygen and carbon species H$_{2}$O and CO, but has a much greater effect on minor constituents CH$_{4}$ and CO$_{2}$, with CH$_{4}$ being present at high C/O ratios and CO$_{2}$ at low C/O ratios. The higher C/O ratio model therefore exhibits more absorption in the 3.6 micron bandpass due to the increased presence of CH$_{4}$, but the relative insignificance of both CH$_{4}$ and CO$_{2}$ under these conditions means that varying the C/O ratio over this range has much less of an effect on the spectral slope than changing the atmospheric metallicity.

\begin{figure}[h!]
\includegraphics[height=.45\textwidth]{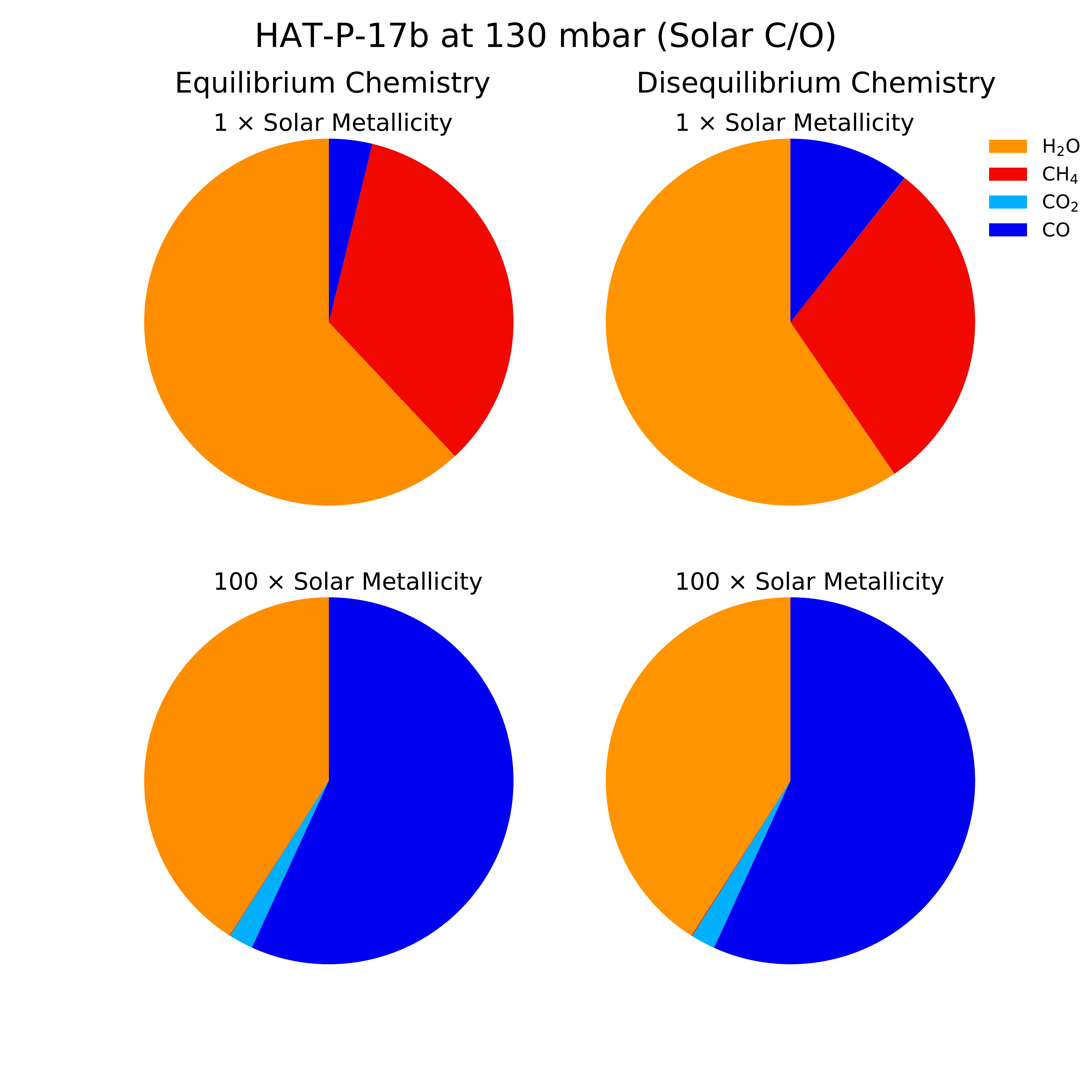}
\caption{Comparison of chemical abundances of H$_{2}$O, CH$_{4}$, CO$_{2}$, and CO from the equilibrium chemical models (left) and disequilibrium chemical models (right) for 1$\times$ solar metallicity models (top) and 100$\times$ solar metallicity models (bottom) all with solar C/O ratios (C/O=0.6) at 130 mbar for our coolest planet, HAT-P-17b.}
\label{fig:H17_pie}   
\end{figure}

\begin{figure}[h!]
\includegraphics[height=.45\textwidth]{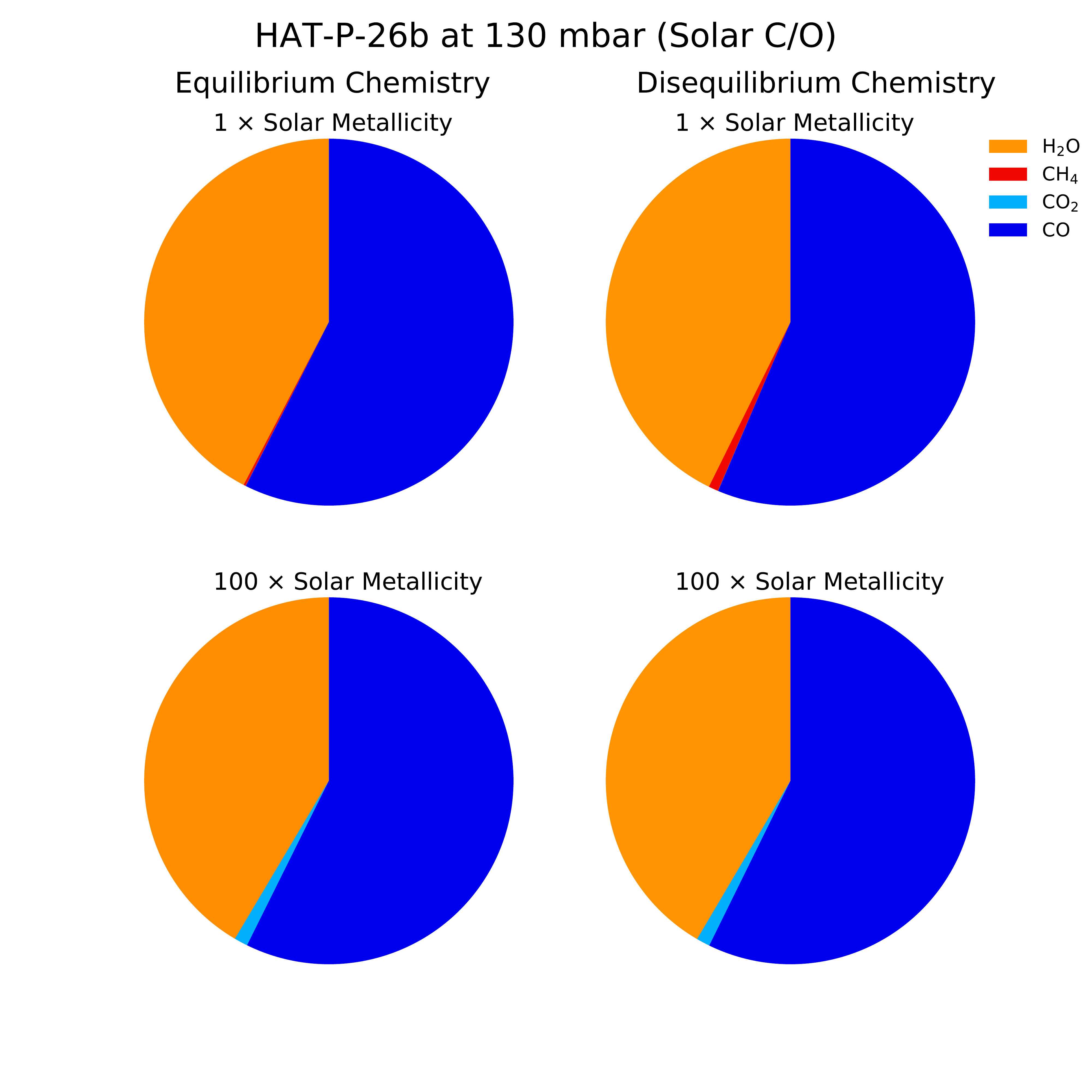}
\caption{Comparison of the relative abundances of H$_{2}$O, CH$_{4}$, CO$_{2}$, and CO from the equilibrium chemical models (left) and disequilibrium chemical models (right) for 1$\times$ solar metallicity models (top) and 100$\times$ solar metallicity models (bottom) all with solar C/O ratios (C/O=0.6) at 130 mbar for our hottest planet, HAT-P-26b.}
\label{fig:H26_pie}   
\end{figure}

For cooler planets where CH$_{4}$ is expected to be the dominant carbon-bearing constituent at solar C/O ratios, extreme changes in C/O ratio can lead to significant changes in atmospheric chemistry. As the C/O ratio is reduced, these cooler planets will eventually experience a transition where the carbon shifts from CH$_{4}$ to CO$_{2}$ dominated carbon chemistries, with a corresponding major shift in spectral slope. Warmer CO-dominated planets can also transition to a different chemical regime when the C/O ratio becomes large enough that CH$_{4}$ becomes a major reservoir of carbon (see Figure~\ref{fig:H17_pie} and Figure~\ref{fig:H26_pie}). The degree of the degeneracy between the atmospheric metallicity and the C/O ratio will be, at least in part, governed by the equilibrium temperature of the planet. In planets that are cool enough to have methane, metallicity will be the primary driver for the spectral slope (as shown by the fact that the C/O ratio does not greatly impact the spectral slope of HAT-P-17b in Figure ~\ref{fig:H17_contour}). For warmer planets without significant methane, the spectral slope will be more strongly influenced by variations in the C/O ratio as shown in Figure ~\ref{fig:H26_contour} and to a lesser degree Figure ~\ref{fig:W69_contour}.

As shown in Figures~\ref{fig:H17_pie} and \ref{fig:H26_pie}, as well as Figures \ref{fig:h17_z1} -~\ref{fig:h26_z100}, the inclusion of disequilibrium chemical processes does not appreciably change this picture (e.g., \citealt{Moses2013Compositional436b}). The disequilibrium models have slightly less CH$_4$ and more CO and CO$_2$ in their upper atmospheres as compared to the equilibrium models, but this is a relatively minor shift compared to the change in chemistry as we vary the atmospheric metallicity and C/O ratio. We therefore conclude that under the thermal conditions relevant to the planets in our sample, variations in atmospheric metallicity are the dominant factor shaping the 3.6-4.5 $\mu$m spectral slopes, unless the atmosphere is highly enriched (C/O$>1-1.5$) or depleted (C/O$<0.1$) in carbon compared to oxygen. Although a more careful consideration of the 3D coupled chemistry and dynamics may alter this picture (e.g., \citealt{Cooper2005}; \citealt{Bordwell2018}; \citealt{Drummond2018}; \citealt{Mendonca2018}; \citealt{Steinrueck2018}), these effects are expected to be relatively minor, as the composition in the infrared photosphere tends to be homogenized to those of the warmest dayside regions.

\begin{centering}
\begin{figure}[h!]
\includegraphics[width=0.47\textwidth]{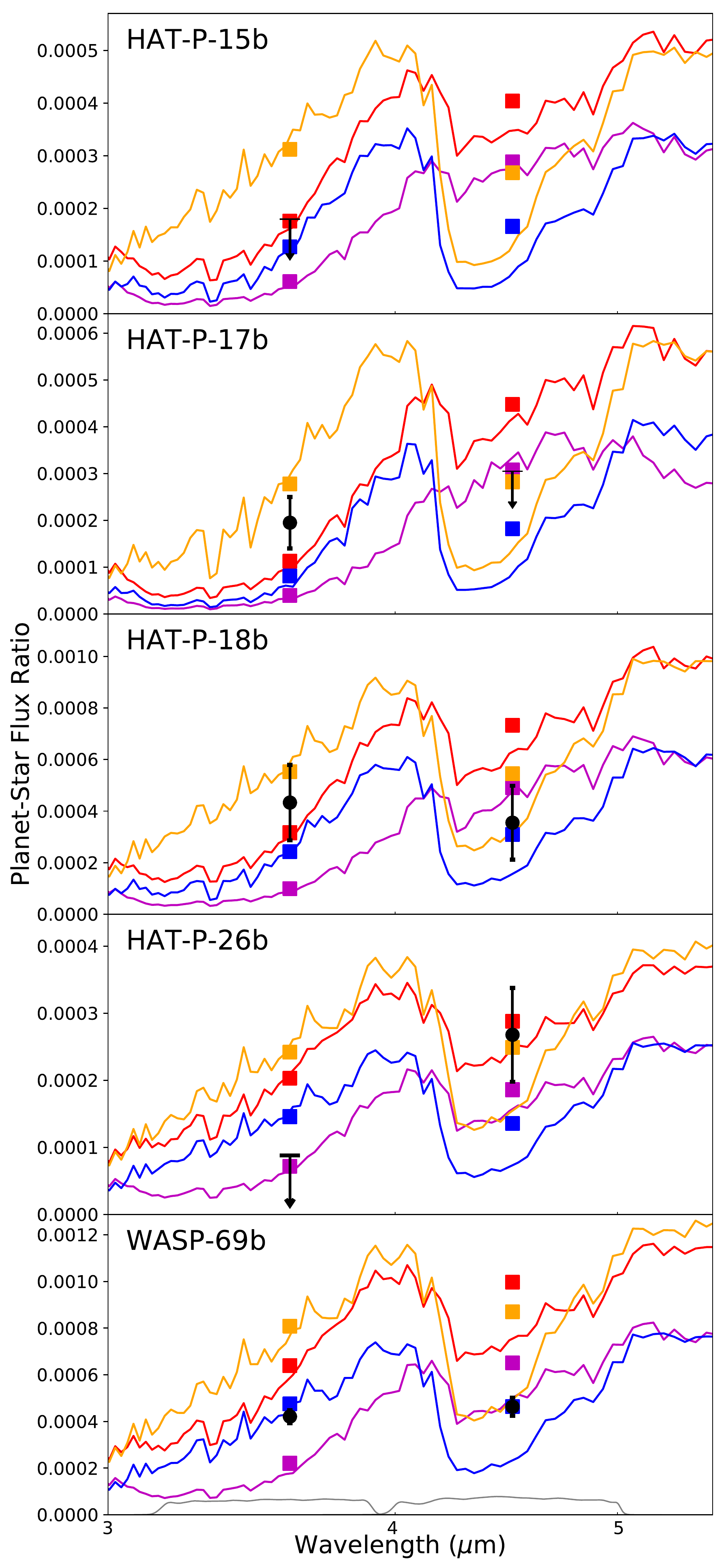}
\caption{Planet-star flux ratios as a function of wavelength from 1D atmosphere models for all five planets. We show 1$\times$ solar and 100$\times$ solar metallicity models with either efficient (purple and blue, respectively) or inefficient (red and orange, respectively) redistribution of energy to the planet's nightside. Our measured eclipse depths for each planet are shown as black circles, and we plot the corresponding band-integrated flux values from the models as filled squares.  For HAT-P-15b (3.6 $\mu$m only; see Section \ref{sec:fits} for more details), HAT-P-17b (4.5 $\mu$m only), and HAT-P-26b (3.6 $\mu$m only) we show the $2\sigma$ upper limit on the eclipse depth.  We also overplot the IRAC 3.6 and 4.5 $\mu$m response functions in gray in the bottom panel for comparison.}
\label{fig:models}   
\end{figure}
\end{centering}

\subsubsection{Model Comparison for Individual Planets}
We next compare our measured secondary eclipse depths for each individual planet to a grid of four models, including atmospheric metallicities of either 1$\times$ or 100$\times$ solar and either full recirculation (i.e., complete redistribution of heat to the planet's nightside) or dayside-only recirculation (i.e., redistribution of heat limited to the dayside hemisphere alone). This set of four models represents a reasonably compact sampling of the possible parameter space, with redistribution efficiency allowing us to make each model globally hotter or cooler while varying atmospheric metallicity serves as a simplified proxy for changes in atmospheric chemistry that can affect the spectral slope. The resulting models are shown in Figure ~\ref{fig:models}.  Although we include the 3.6 $\mu$m depth for HAT-P-15b for completeness, we do not detect an eclipse in either bandpass, and therefore refrain from any further discussion of the implications of the eclipse depths for our understanding of this planet's atmosphere. 

As in \citet{Kammer2015SPITZERSPECTRA}, model-data comparisons for all four planets  with detected eclipses strongly prefer models with efficient circulation between the day- and nightside hemispheres.  We find that the HAT-P-26b eclipses are best matched by the 1$\times$ solar metallicity model, in good agreement with the constraints from transmission spectroscopy presented in \citet{Wakeford2017HeavyAbundance}. The WASP-69b eclipses are well matched by the 100$\times$ solar metallicity model. Our constraints for HAT-P-17b and HAT-P-18b are somewhat weaker, but still appear to modestly favor the 100$\times$ solar metallicity model over the 1$\times$ solar model. 

\subsection{Model-independent Trends in Atmospheric Composition}  \label{sec:Tbright}
We next consider the same model-independent metric used in \citet{Kammer2015SPITZERSPECTRA} to search for trends in atmospheric composition.  This metric is defined as the ratio of the measured brightness temperatures in the 4.5-3.6 $\mu$m bandpasses, and should be effectively independent of the planet-star radius ratio and equilibrium planet temperature.  We expect variations in this ratio to instead reflect the relative strength of CH$_{4}$ (3.6 $\mu$m) versus CO and CO$_2$ (4.5 $\mu$m) absorption features in the atmospheres of these planets.  We use this ratio to search for empirical correlations with other parameters of interest including the planet's mass, bulk metallicity, and host star metallicity.

\begin{centering}
\begin{deluxetable*}{lccccccc}[t!]
\tabletypesize{\scriptsize}
\tablecaption{Brightness Temperature Ratios}
\tablewidth{0pt}
\tablehead{
\colhead{Planet}& \colhead{Mass (M$_{\tt Jup}$)} & \colhead{Radius (R$_{\tt Jup}$)}  &\colhead{ T$_{\tt eq}$ (K)}  &\colhead{T$_\textsubscript{Bright}$ Ratio} &\colhead{Bulk Z$_{planet}$} &\colhead{[Fe/H]$_{*}$} &\colhead{Ref} }
\startdata
GJ 436b &0.07 $\pm$ 0.01 &0.37 $\pm$ 0.02&669 $\pm$ 22&$<$ 0.72 &0.83 $\pm$ 0.02 &0.05 $\pm$ 0.141 &1,2,3,4\\
GJ 3470b &0.0432 $\pm$ 0.0051&0.346 $\pm$ 0.029&604 $\pm$ 98 &$<$ 1.00 & 0.74 $\pm$ 0.37&0.27 $\pm$ 0.11 &5,6,7\\
HAT-P-12b &0.21 $\pm$ 0.01  &0.96$^{+0.03}_{-0.02}$&963 $\pm$ 16& 1.06$_{-0.10}^{+ 0.08}$ &0.32 $\pm$ 0.03&-0.26 $\pm$ 0.06 &1,8,9,10 \\
HAT-P-17b & 0.537 $\pm$ 0.017&1.010 $\pm$ 0.029 &791$\pm$17 & $<$0.87&0.13 $\pm$ 0.04& 0.05 $\pm$ 0.03&1,10,11\\
HAT-P-18b & 0.200 $\pm$ 0.019& 0.995 $\pm$ 0.052&822 $\pm$ 22&0.78$_{-0.09}^{+ 0.08}$ &0.24 $\pm$ 0.05 & 0.10 $\pm$ 0.08&1,10,11\\
HAT-P-19b & 0.292 $\pm$ 0.018& 1.132 $\pm$ 0.072&1010 $\pm$ 42& 0.84 $\pm {0.06}$ &0.22 $\pm$ 0.05&0.29 $\pm$ 0.06 & 1,10,12,13\\
HAT-P-20b & 7.25 $\pm$ 0.19&0.87 $\pm$ 0.03&970 $\pm$ 23&1.00$_{-0.04}^{+ 0.03}$ &0.28 $\pm$ 0.05&0.12 $\pm$ 0.15 &1,10,14,15\\
HAT-P-26b & 0.059 $\pm$ 0.007 & 0.565$_{-0.032}^{+0.072}$& 1028 $\pm$ 21& $>$ 1.15&0.66 $\pm$ 0.03&0.01 $\pm$ 0.04 &1,10,11 \\
WASP-8b &2.24$^{+0.08}_{-0.09}$&1.04$^{+0.01}_{-0.05}$ &948 $\pm$ 22 &  0.73 $\pm {0.06}$&0.11 $\pm$ 0.04& 0.29 $\pm$ 0.03 &1,10,16\\
WASP-10b &3.14 $\pm$ 0.27&1.039 $^{+0.043}_{-0.049}$&972 $\pm$ 31&  0.94$\pm$ 0.03&0.12 $\pm$ 0.02&0.04 $\pm$ 0.05 &1,2,10,13\\
WASP-67b &0.406 $\pm$ 0.035&1.091 $\pm$ 0.046&1003 $\pm$ 20& $>$0.97 &0.2 $\pm$ 0.06&0.18 $\pm$ 0.06 &1,10,13,17\\
WASP-69b & 0.250 $\pm$ 0.023& 1.057 $\pm$ 0.047& 961 $\pm$ 20& 0.85 $\pm$ 0.02 &0.21 $\pm$ 0.04&0.30 $\pm$ 0.06 &1,10,11\\
WASP-80b &0.54 $\pm$ 0.04&1.00 $\pm$ 0.03&825 $\pm$ 19& 0.99 $\pm$ 0.10 &0.17 $\pm$ 0.04& 0.13 $\pm$ 0.11&1,7,9,18\\
\enddata 
\tablenotetext{}{\textbf{References.} (1) \citet{Thorngren2019}, (2) T$_\texttt{eq}$ from \citet{Southworth2011HomogeneousCurves}, (3) \citet{Morley2017ForwardClouds}, (4) [Fe/H] from \citet{Rojas-Ayala2012MetallicityMdwarfs}, (5) \citet{Biddle2014}, (6) \citet{Benneke2019}, (7) [Fe/H] from \citet{Terrien2015ADwarfs}, (8) \citet{Hartman2009}, (9) Wong et al. (2019, in preperation), (10) [Fe/H] from \citet{Santos2013SWEET-Cat:ExoplanETs}, \citet{Sousa2018}, (11) This work, (12) \citet{Hartman2011HAT-P-18bStars}, (13) \citet{Kammer2015SPITZERSPECTRA}, (14) \citet{Bakos2011}, (15) \citet{Deming2015SpitzerDecorrelation}, (16) \citet{Cubillos2013WASP-8b:Spitzer}, (17) \cite{mancini2014}, (18) \citet{Triaud2015WASP-80bRange}}
\label{table:systems}
\end{deluxetable*}
\end{centering}

We note that the presence or absence of a temperature inversion might also alter a planet's relative brightness in these two bands in a way that mimics the shift from a CH$_{4}$-dominated to CO- and CO$_2$-dominated carbon chemistry.  Observations of the broader sample of hot Jupiters suggests that temperature inversions are only found in the atmospheres of the most highly irradiated planets and are most likely caused by the presence of gas phase TiO (e.g., \citealt{Evans2017AnStratosphere}; \citealt{Nugroho2017High-ResolutionWASP-33b}; \citealt{Sedaghati2017DetectionJupiter}; \citealt{Sheppard2017EvidenceWASP-18b}).  The planets in this study are too cool for TiO and VO to remain in the gas phase, and there is currently no evidence for temperature inversions in the atmospheres of planets at these temperatures.

Two of the Neptune-mass planets in our sample, GJ 436b and HAT-P-26b, have previously published constraints on their atmospheric metallicities and C/O ratios. \citet{Morley2017ForwardClouds} re-examined all of the available secondary eclipse data for GJ 436b, which has a low 4.5-3.6 $\mu$m brightness temperature ratio.  They concluded that these observations were consistent with absorption features from water, carbon monoxide, and carbon dioxide corresponding to an atmospheric metallicity greater than 200$\times$ solar and a C/O ratio consistent with solar. Similarly, transmission spectroscopy for HAT-P-26b from \citet{Wakeford2017HeavyAbundance} constrains its atmospheric metallicity to $0.8-26\times$ solar ($1\sigma$); this is consistent with our measurement of a relatively high 4.5-3.6 $\mu$m brightness temperature ratio.  We therefore conclude that for these two planets, differences in atmospheric metallicity are likely the primary factor driving the observed difference in $3.6-4.5$ $\mu$m spectral slope, in good agreement with our predictions based on the model grids in Section \ref{sec:chemistry}.

\begin{figure*}
\includegraphics[width=1.\textwidth]{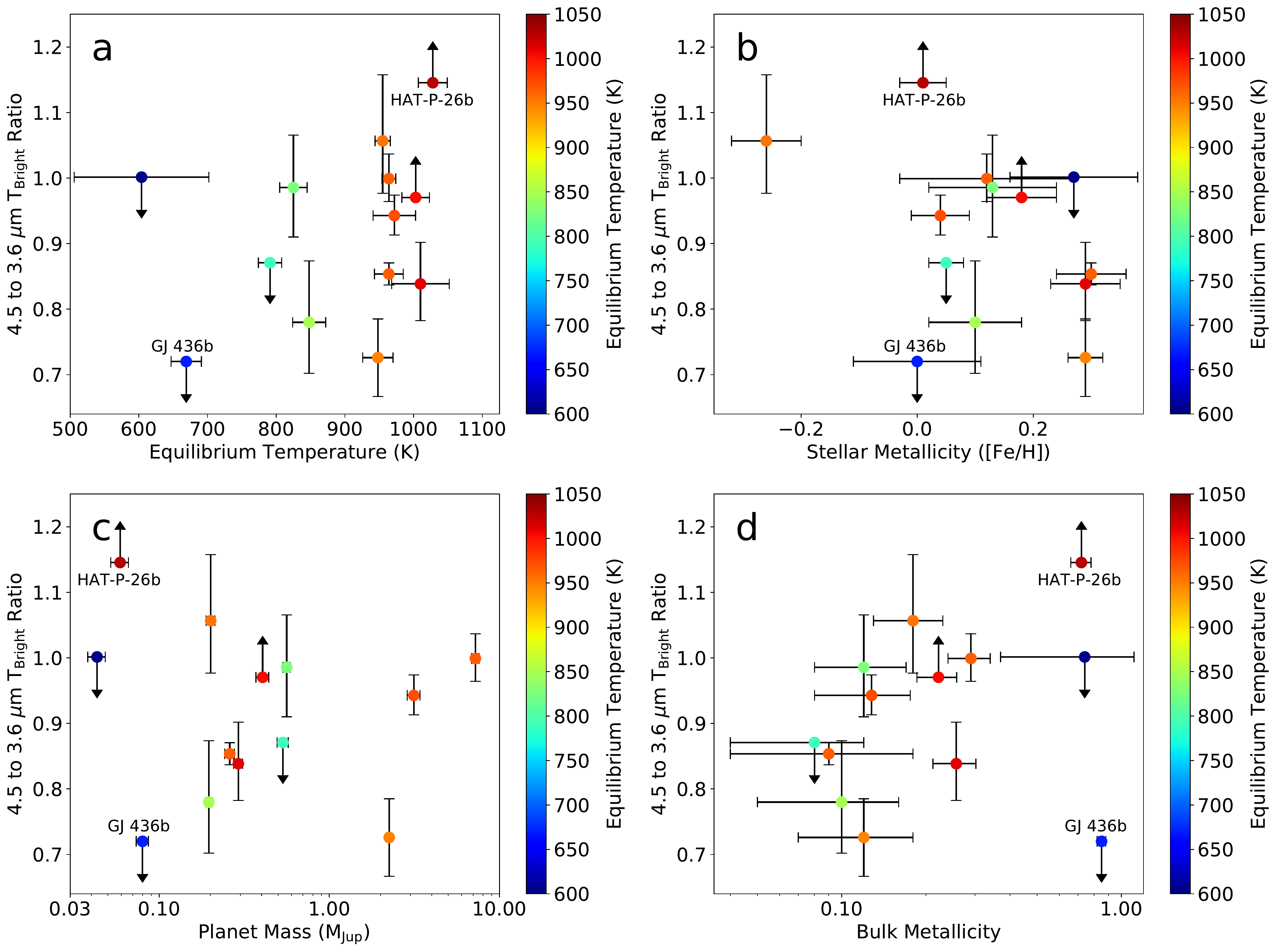}
\caption{Measured 4.5-3.6 $\mu$m brightness temperature ratio as a function of (a) equilibrium temperature, (b) stellar metallicity, (c) planetary mass, and (d) bulk metallicity for all planets with published \emph{Spitzer} eclipse depths and equilibrium temperatures less than 1100 K. We show four planets from this study (HAT-P-17b, HAT-P-18b, HAT-P-26b, and WASP-69b) as well as nine previously published planets (see Table ~\ref{table:systems} for the full list of planets and corresponding references). For planets with no eclipse detected in the 3.6 (4.5) $\mu$m bands we plot 2$\sigma$ lower (upper) limits, respectively.}
\label{fig:ratio}
\end{figure*}

We use the measured spectral slopes for our sample of planets to search for trends in spectral shape across our sample of short-period gas giant planets.  If these planets have broadly solar C/O ratios and follow the same trend of increasing atmospheric metallicity with decreasing planet mass that we see for the solar system gas giants, we would expect to see a rising trend in the measured brightness temperature ratios with increasing planet mass.  In Figure~\ref{fig:ratio}{\color{blue}c} we plot this brightness temperature ratio as a function of planet mass for the four planets with measured eclipses described herein as well as other planets with temperatures less than 1100 K and published secondary eclipse detections (see Table~\ref{table:systems}). As before, we see that planets with eccentric orbits (GJ 436b, HAT-P-17b, and WASP-8b) appear to have brightness temperature ratios that are systematically lower than those of planets on circular orbits with the same mass. Other than this apparent clustering of eccentric planets, there appears to be no obvious correlation between atmospheric composition and planetary mass for planets in this temperature regime. 

We also check to see if the atmospheric compositions of these planets are correlated with their bulk metallicities. We take published bulk metallicity values from \citet{Thorngren2019}, which used a 1D planetary model with an inert rock-ice core and a convective H/He-rock-ice envelope. We calculate new bulk metallicity values for GJ 3470b, which was not included in this study, using the same method (see Table \ref{table:systems}). We find evidence for a correlation between atmospheric metallicity and bulk metallicity (Figure ~\ref{fig:ratio}{\color{blue}d}), where planets with higher bulk metallicities have on average slightly higher 4.5-3.6 $\mu$m brightness temperature ratios. 
We evaluate the significance of the proposed trend in brightness temperature ratio versus bulk metallicity using a Monte Carlo simulation where we create a series of simulated data sets by sampling from the posterior probability distributions for each point. For the brightness temperature ratios, we assume that the reported uncertainties in the 3.6 and 4.5 $\mu$m brightness temperatures are reasonably well approximated by  Gaussian distributions.  We then draw 10$^{6}$ samples from these two distributions for each planet. 

For planets with a nondetection in one band, we assume that the brightness temperature in the band with the non-detection follows a Gaussian distribution with a standard deviation of 100 K centered at the assumed median of the distribution as determined from the brightness temperature corresponding to the eclipse depth of the reported 2$\sigma$ upper limit (i.e., the center of the distribution is the brightness temperature corresponding to the 2$\sigma$ upper limit - 200 K); this effectively excludes solutions in which the brightness temperature for the nondetection is unphysically low (i.e., 98\% of all samples are limited to temperatures within 400 K of the 2$\sigma$ upper limit). Lastly, we assume that the bulk metallicities are also well approximated by Gaussian distributions and sample from those as well.

If we exclude the three Neptune-mass planets (GJ 436b, GJ 3470b, and HAT-P-26b), we do find relatively weak (with positively sloped lines preferred over flat or negatively sloped lines at the 1.4$\sigma$ level) evidence for a linear trend with bulk metallicity. If it can be substantiated with additional measurements, this correlation would be somewhat surprising as it would suggest that gas giant planets with high bulk metallicities may have lower atmospheric metallicities.  This would be the opposite of the observed trend for the solar system gas giants, in which planets with higher bulk metallicities also have higher atmospheric metallicities.  

Intriguingly, we do see tentative evidence for a trend in atmospheric composition with stellar metallicity (Figure~\ref{fig:ratio}{\color{blue}b}). This trend is not entirely surprising given that we would expect metal-rich stars to have correspondingly metal-rich disks. However, the metallicities of the host stars in our sample only vary between $-0.26$ and $+0.29$, corresponding to a relatively small 3.5$\times$ change in bulk disk metallicity.  In contrast, we would need to vary the atmospheric metallicities of these planets by more than two orders of magnitude to reproduce the observed change in 4.5-3.6 $\mu$m brightness temperature ratios.
If real, this trend suggests that gas giant planets in metal-rich disks incorporate disproportionately more solids into their atmospheres than planets in metal-poor disks.  If we assume that these planets formed in the inner disk, this might be explained by scenarios in which gas drag causes approximately centimeter-sized particles to migrate inward, increasing the concentration of solids in this region \citep[e.g.,][]{Johansen2014TheProcess}.

\begin{centering}
\begin{figure}[h!]
\includegraphics[width=.45\textwidth]{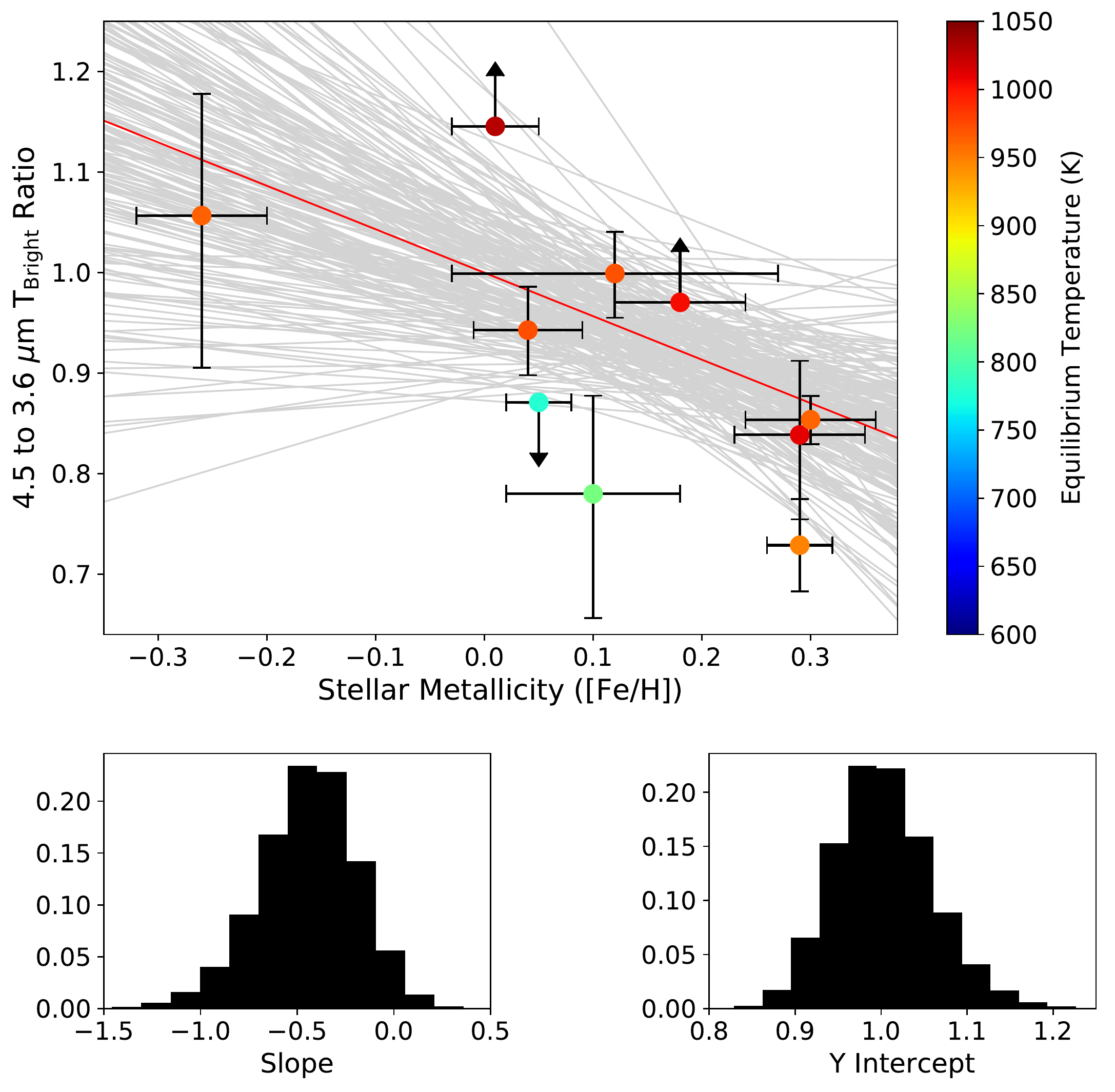}
\caption{Measured 4.5 $\mu$m to 3.6 $\mu$m brightness temperature ratio as a function of stellar metallicity. We create 10$^6$ simulated data sets by sampling from posterior probability distributions for the reported brightness temperatures and stellar metallicities, and show a random subset of the resulting distribution of linear fits as gray lines.  We indicate the median linear solution in red, and plot the corresponding distributions in $y$ intercept and slope below.}
\label{fig:mc_stellarZ}
\end{figure}
\end{centering}

We evaluate the significance of this proposed trend in the same manner as before (i.e., by drawing 10$^{6}$ samples from the distributions for the brightness temperatures as described above) and assuming that the reported stellar metallicities are well approximated by Gaussian distributions.  We then fit the resulting 10$^{6}$ simulated data sets with a linear function and plot the corresponding posterior probability distributions for the slope and $y$ intercept of this line (Figure ~\ref{fig:mc_stellarZ}). We find conflicting literature values for the metallicity of the M-dwarf stars in our sample (GJ 436, GJ 3470, and WASP-80), many of which also have relatively large uncertainties.  We therefore opted to exclude these M-dwarf hosts when evaluating the significance of the proposed trend (\citealt{Rojas-Ayala2012MetallicityMdwarfs}; \citealt{Terrien2012ANDwarfs}; \citealt{Lindgren2017MetallicityTemperature}).  We find that negative slopes are preferred, but our data are still consistent with a positive sloped or flat line at the $1.9\sigma$ level.

We also consider a scenario in which the points shown in Figure ~\ref{fig:mc_stellarZ} do not follow a smooth trend, but rather comprise two distinct groups of planets-- those with high atmospheric metallicities and those with low atmospheric metallicities.  We follow the same approach as \citet{Schlaufman2018EvidenceFormation} and try a hierarchical clustering algorithm, a k-means clustering algorithm, and a Gaussian-model clustering algorithm.  All three algorithms are available as part of the \texttt{scikit-learn} package \citep{Pedregosa2011Scikit-learn:Python}. We find that these three methods were inconclusive as to the membership of the planets within the two groups, indicating that these measurements are consistent with a single population.

\section{Conclusions}\label{sec:conclusions}

We present new secondary eclipse depth measurements in the 3.6 and 4.5 $\mu$m \emph{Spitzer} bands for HAT-P-17b, HAT-P-18b, HAT-P-26b, and WASP-69b and place upper limits on the eclipse depths of HAT-P-15b.  Our measured times of secondary eclipse for HAT-P-18b, HAT-P-26b, and WASP-69b are consistent with circular orbits, and we confirm the non-zero eccentricity of HAT-P-17b. We compare our measured eclipse depths with 1D radiative-convective models for each planet.  For HAT-P-26b, which was the only planet with a well-constrained atmospheric metallicity from transmission spectroscopy, our data are in good agreement with the low atmospheric metallicity reported in \citet{Wakeford2017HeavyAbundance}. We find no evidence for a correlation between atmospheric composition and planetary mass. However, we do find a suggestive $1.9\sigma$ trend in atmospheric composition as a function of stellar metallicity. While the existence of a correlation between planetary atmospheric composition and stellar metallicity would not be surprising, the strength of the observed trend implies that short-period gas giant planets orbiting metal-rich stars may have atmospheric metallicities that are significantly higher than the bulk metallicity of the disk. However, our ability to fully understand this possible trend is limited by the availability of precise stellar metallicity measurements. 

\begin{centering}
\begin{figure}
\includegraphics[width=.45\textwidth]{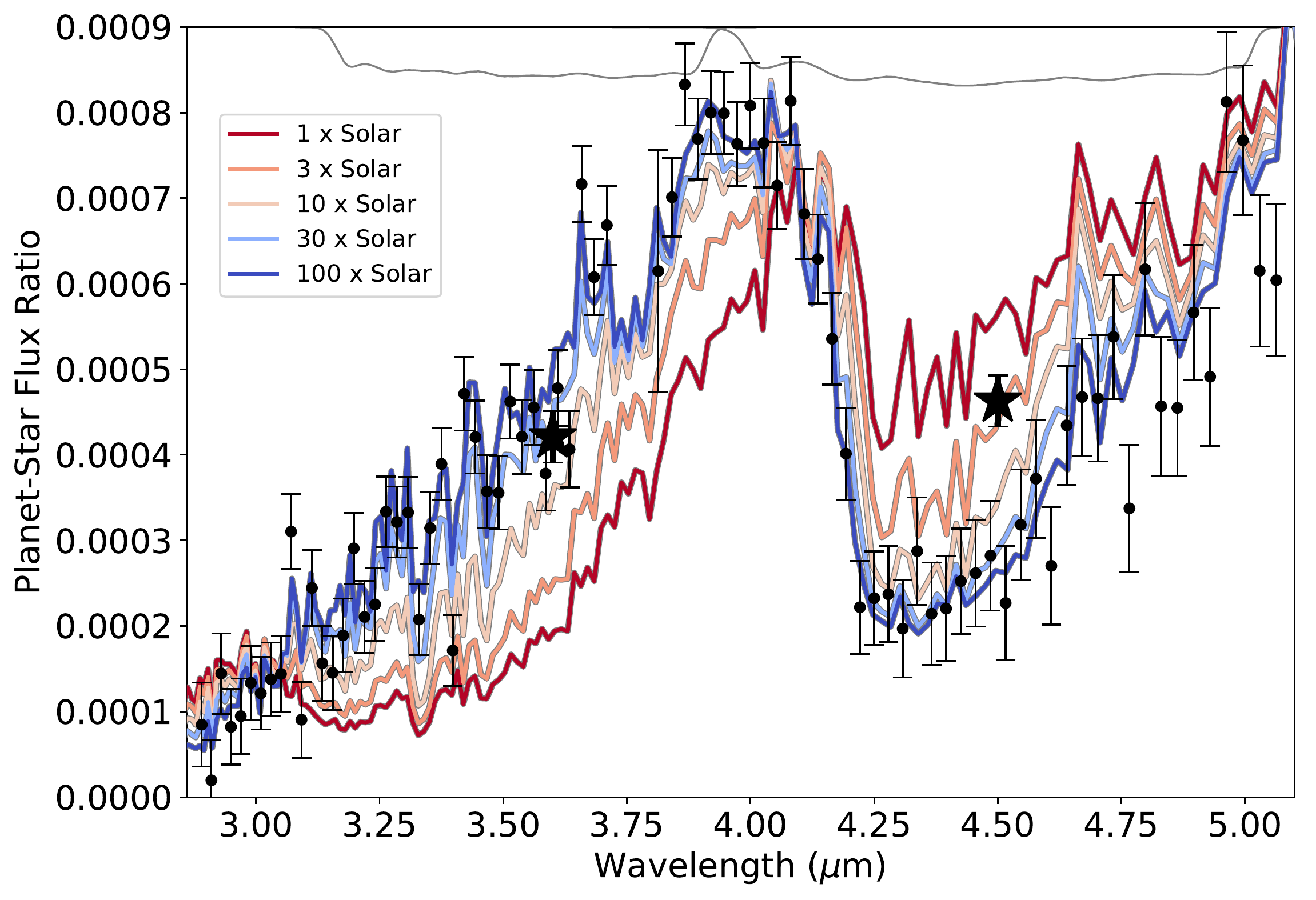}
\caption{Simulated \textit{JWST} observations of a single secondary eclipse observation of WASP-69b using the NIRSpec G395 grism (black points). We use our best-fit 100$\times$ solar metallicity model for this simulation and overplot models with 1$\times$, 3$\times$, 10$\times$ and 30$\times$ solar metallicities for comparison (all with solar C/O ratios). The IRAC 3.6 and 4.5 $\mu$m response functions are shown in gray and our measured eclipse depths in the two \emph{Spitzer} bands are shown as filled black stars.} 
\label{fig:jwst}
\end{figure}
\end{centering}

Beginning in 2021, the broad infrared wavelength coverage and higher spectral resolution of the \textit{James Webb Space Telescope} (\textit{JWST}) will provide invaluable new insights into the atmospheric compositions of this population of planets (see Figure~\ref{fig:jwst} for the expected \textit{JWST} precision and coverage). In this study we relied on broad photometric bandpasses that span multiple absorption features including water, methane, carbon monoxide, and carbon dioxide.  This necessarily results in degeneracies in the interpretation of these data, including correlations between the abundances of the various molecular species and also with the planet's dayside pressure-temperature profile.  \textit{JWST} will be able to resolve individual molecular bands, therefore avoiding these degeneracies and allowing for robust abundance constraints for these atmospheres.  It is worth noting that these cooler planets are also ideal targets for studies examining trends in bulk metallicity as a function of planet mass (e.g., \citealt{Miller2011TheRevealed}, \citealt{Thorngren2016ThePlanets}), as hotter planets have inflated radii that make it difficult to accurately determine their bulk metallicities (e.g., \citealt{Laughlin2011OnPlanets}, \citealt{Thorngren2018BayesianDissipation}).
We therefore expect that future studies of this cool gas giant population with \textit{JWST} will be able to determine for the first time whether or not the postulated correlation between planet mass, core mass fraction, and atmospheric metallicity is in fact a universal property of all gas giant planets.  This in turn will tell us whether or not the interior and atmospheric compositions of gas giant planets are the inevitable outcome of the core accretion process or instead primarily reflect the diverse formation locations and disk properties of these planets. 

\section{Acknowledgements} 
This work is based on observations made with the \textit{Spitzer Space Telescope}, which is operated by the Jet Propulsion Laboratory, California Institute of Technology under a contract with NASA.  Support for this work was provided by NASA through an award issued by JPL/Caltech.  J.M.D acknowledges that the research leading to these results has received funding from the European Research Council (ERC) under the European Union's Horizon 2020 research and innovation programme (grant agreement no. 679633; Exo-Atmos). J.M. acknowledges support from NASA grant NNX16AC64G.

\bibliographystyle{apj}
\bibliography{Mendeley.bib}

\appendix

\begin{centering}
\begin{figure*}[h!]
\includegraphics[width=\textwidth]{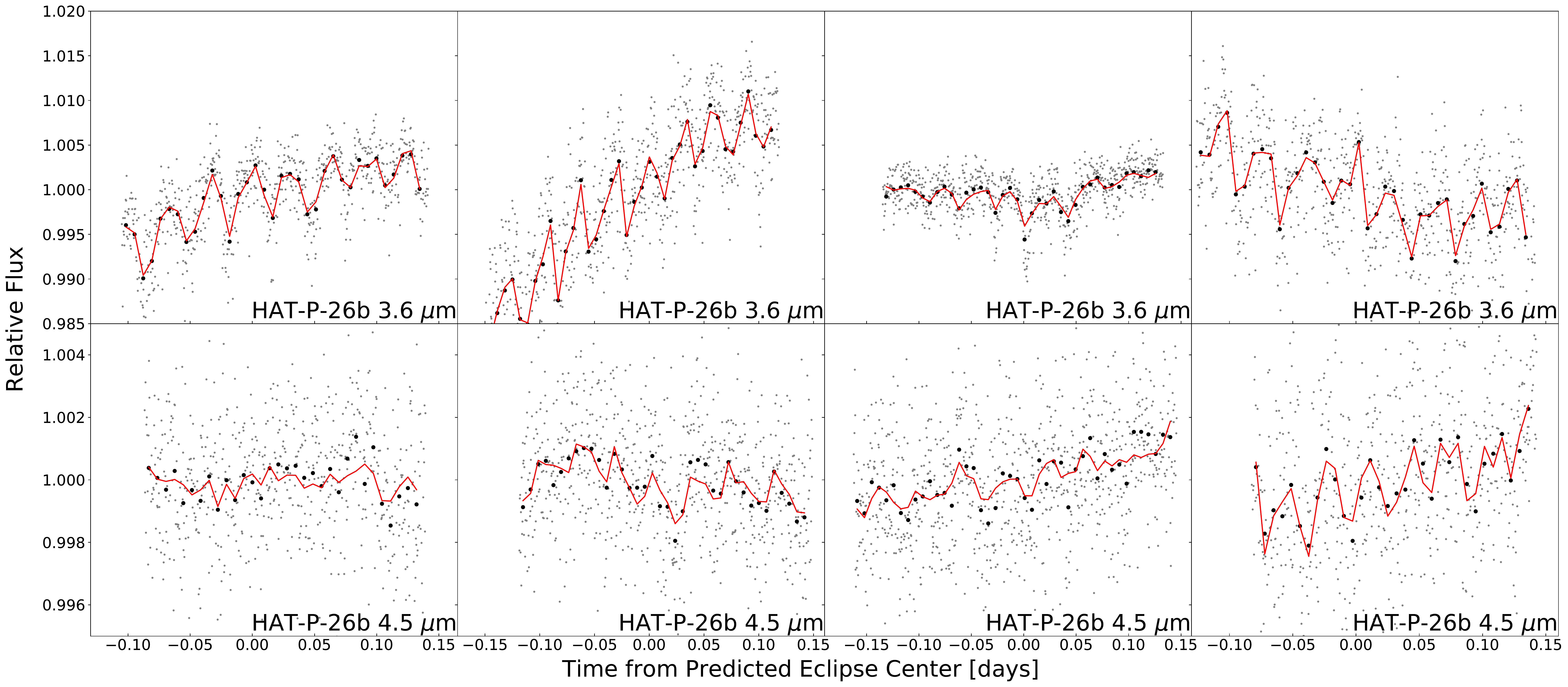}
\caption{Raw \emph{Spitzer} photometry for each visit of HAT-P-26b. The normalized flux binned in ten-minute intervals is shown in black and the thirty-second binned flux is shown in gray. Overplotted is the best-fit instrumental model in red. Observations are shown in chronological order across each row. }
\label{fig:A1}  
\end{figure*}
\end{centering}

\begin{centering}
\begin{figure*}[h!]
\includegraphics[width=\textwidth]{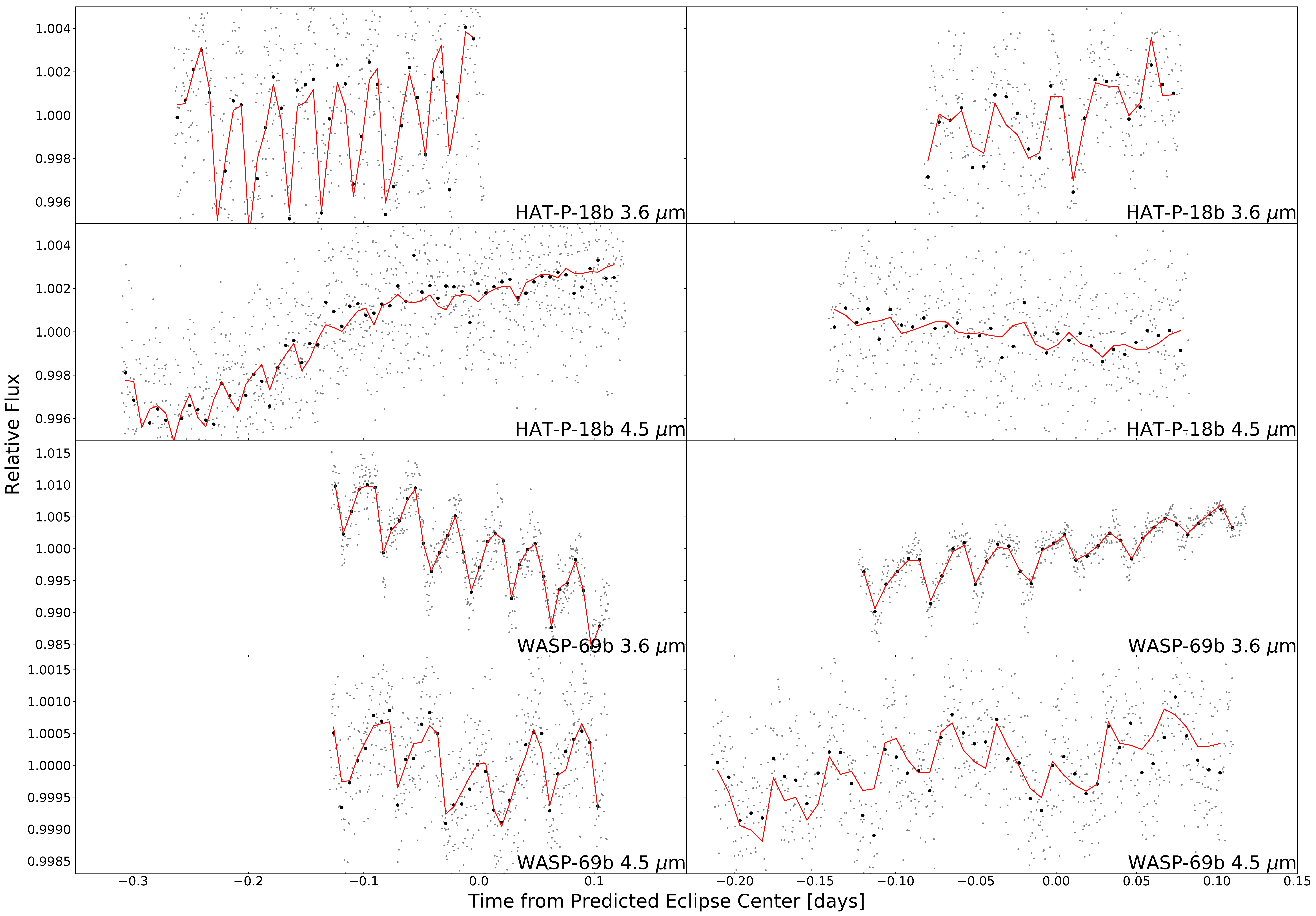}
\caption{Raw \emph{Spitzer} photometry for each visit of HAT-P-18b and WASP-69b. The normalized flux binned in ten-minute intervals  is shown in black and the thirty-second binned flux is shown in gray. Overplotted is the best-fit instrumental model in red. Observations are shown in chronological order across each row.}
\label{fig:A2}  
\end{figure*}
\end{centering}

\begin{centering}
\begin{figure*}[h!]
\includegraphics[width=\textwidth]{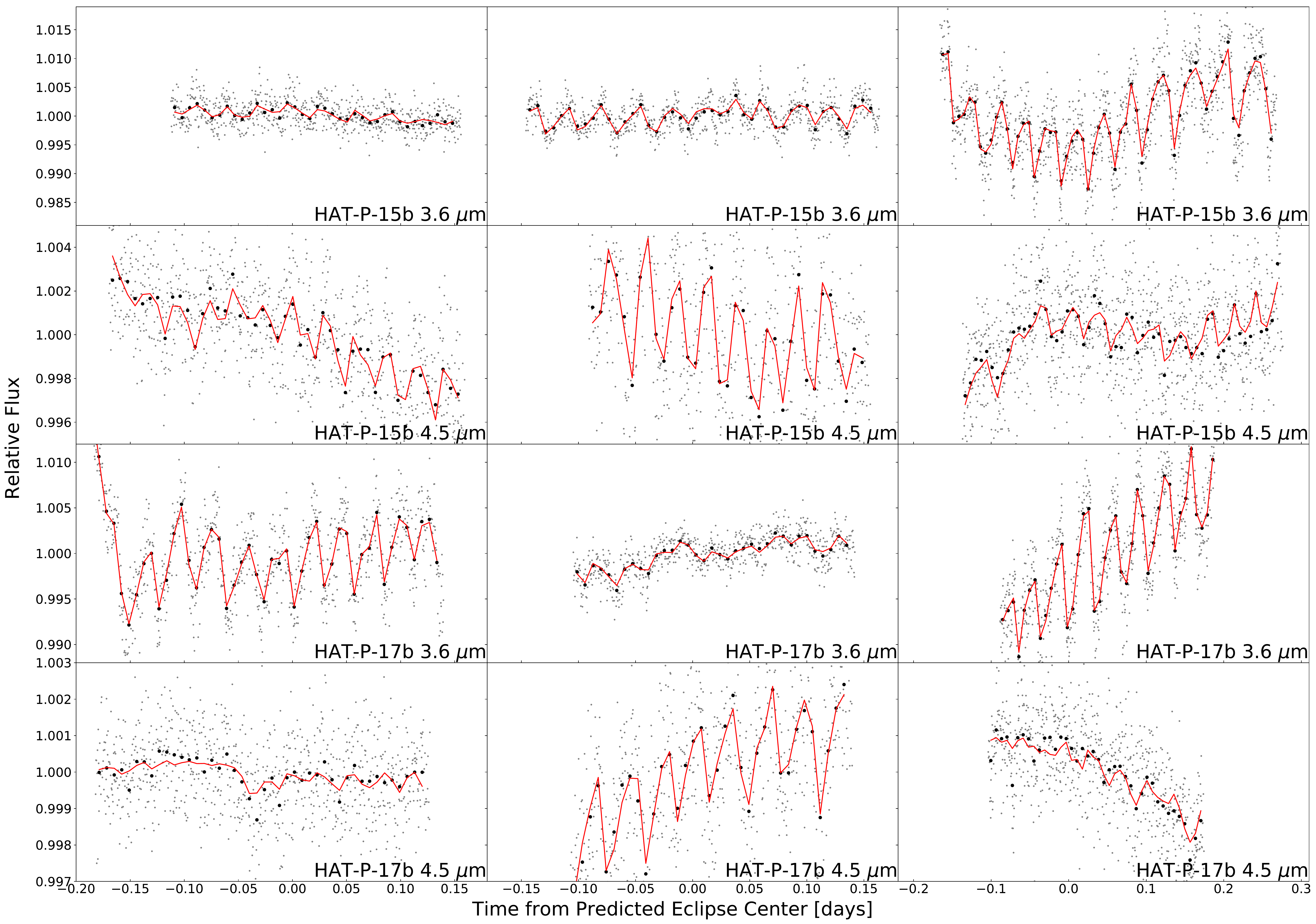}
\caption{Raw \emph{Spitzer} photometry for each visit of HAT-P-15b and HAT-P-17b. The normalized flux binned in ten-minute intervals is shown in black and the thirty-second binned flux is shown in gray. Overplotted is the best-fit instrumental model in red. Observations are shown in chronological order across each row.}
\label{fig:A3}  
\end{figure*}
\end{centering}

\begin{centering}
 \begin{figure*}[t!]
\includegraphics[width=\textwidth]{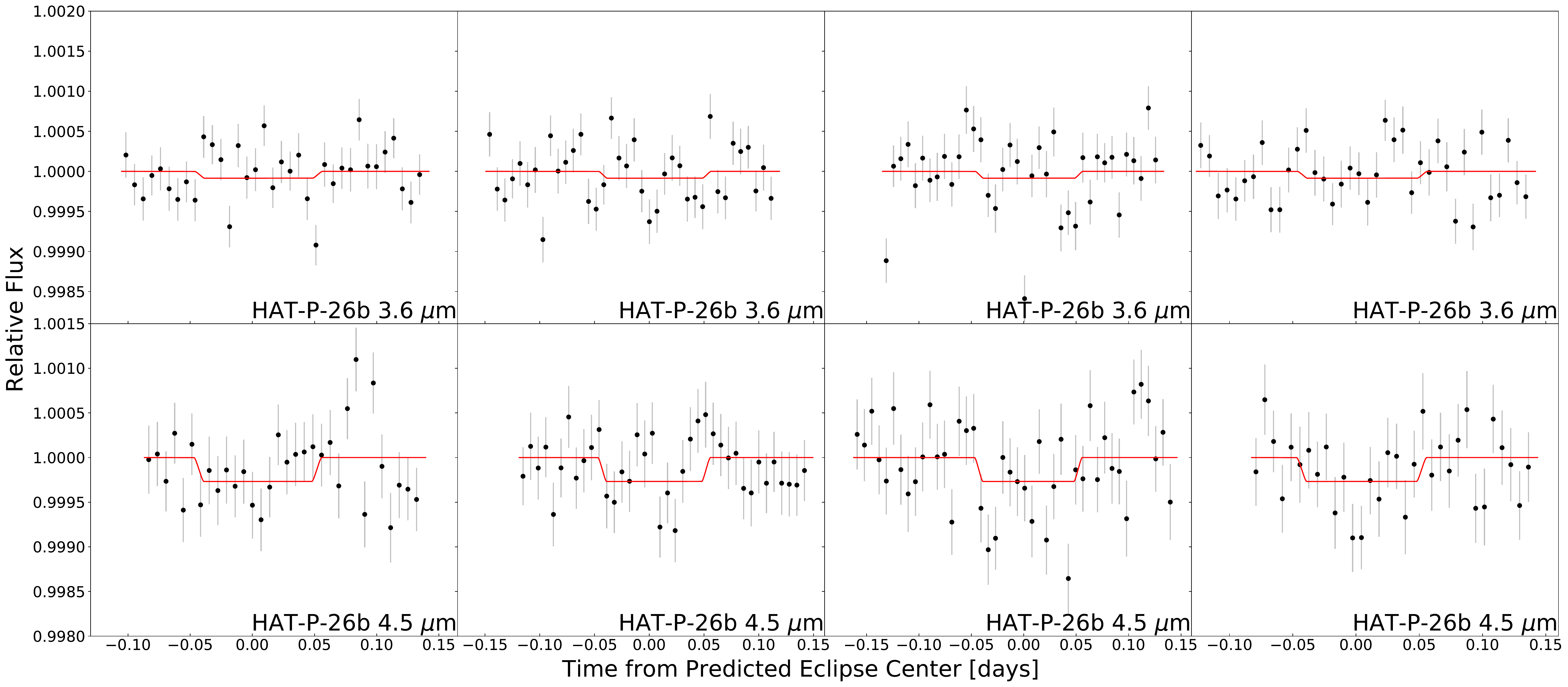}
\caption{Individual light curves for each visit of HAT-P-26b. The ten-minute binned normalized flux in shown in black with error bars showing the standard error of the flux in each bin. The instrumental best-fit parameters are unique to each visit and have been divided out. The red lines are the light curves with the best-fit parameters from the joint fits. Observations are shown in chronological order across each row. The 2$\sigma$ upper limits for the best-fit eclipse depths of the 3.6 $\mu$m visits are shown.}
\label{fig:A4}  
\end{figure*}
\end{centering}

\begin{centering}
 \begin{figure*}[t!]
\includegraphics[width=\textwidth]{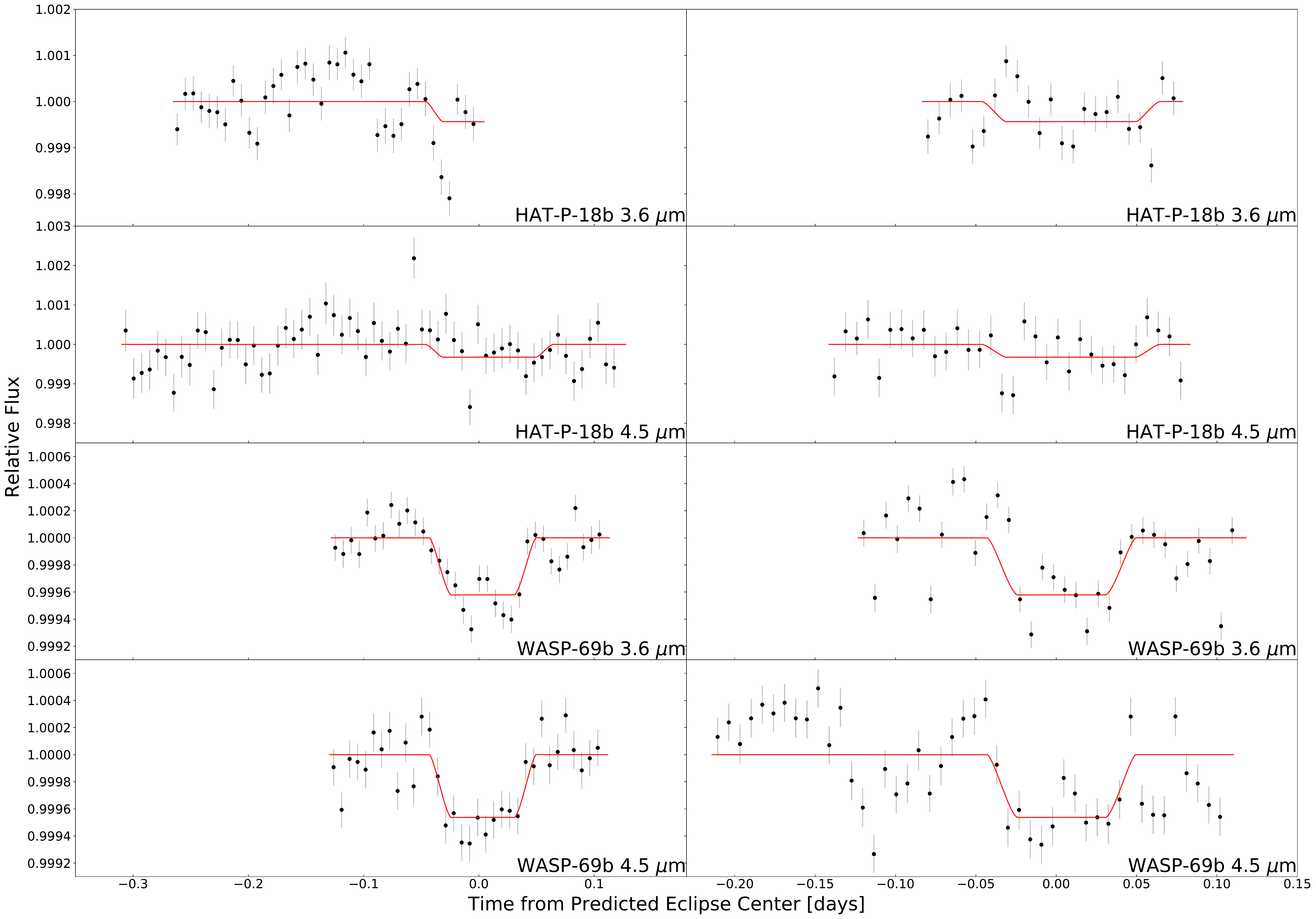}
\caption{Individual light curves for each visit of HAT-P-18b and WASP-69b. The ten-minute binned normalized flux in shown in black with error bars showing the standard error of the flux in each bin. The instrumental best-fit parameters are unique to each visit and have been divided out. The red lines are the light curves with the best-fit parameters from the joint fits. Observations are shown in chronological order across each row.}
\label{fig:A5}  
\end{figure*}
\end{centering}

\begin{centering}
 \begin{figure*}[t!]
\includegraphics[width=\textwidth]{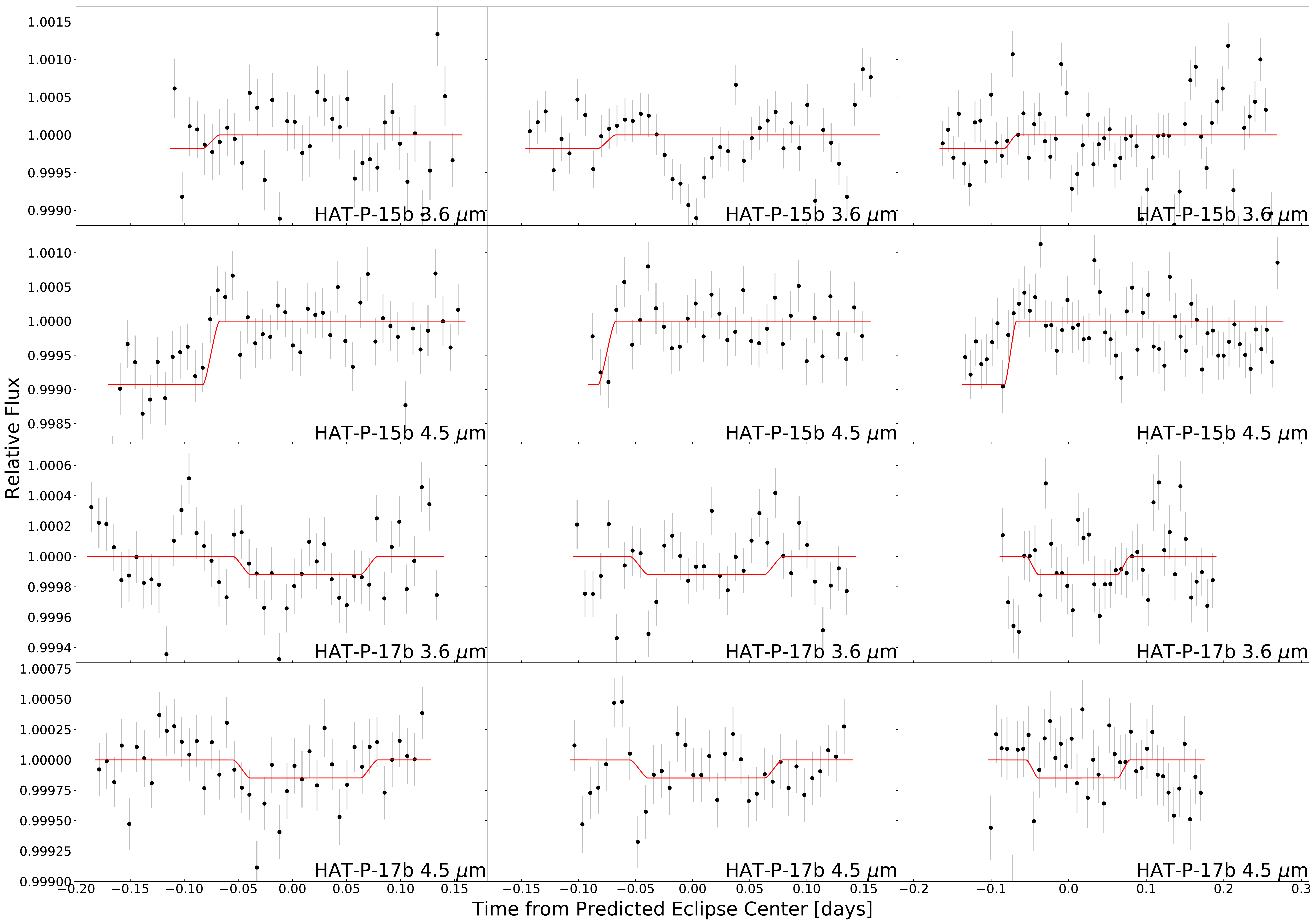}
\caption{Individual light curves for each visit of HAT-P-15b and HAT-P-17b. The ten-minute binned normalized flux in shown in black with error bars showing the standard error of the flux in each bin. The instrumental best-fit parameters are unique to each visit and have been divided out. The red lines are the light curves with the best-fit parameters from the joint fits. Observations are shown in chronological order across each row. The 2$\sigma$ upper limits for the best-fit eclipse depths of all visits of HAT-P-15b (see Section \ref{sec:fits} for more details) and the 4.5 $\mu$m visits of HAT-P-17b are shown.}
\label{fig:A6}  
\end{figure*}
\end{centering}

\begin{centering}
\begin{figure*}[h!]
\includegraphics[width=\textwidth]{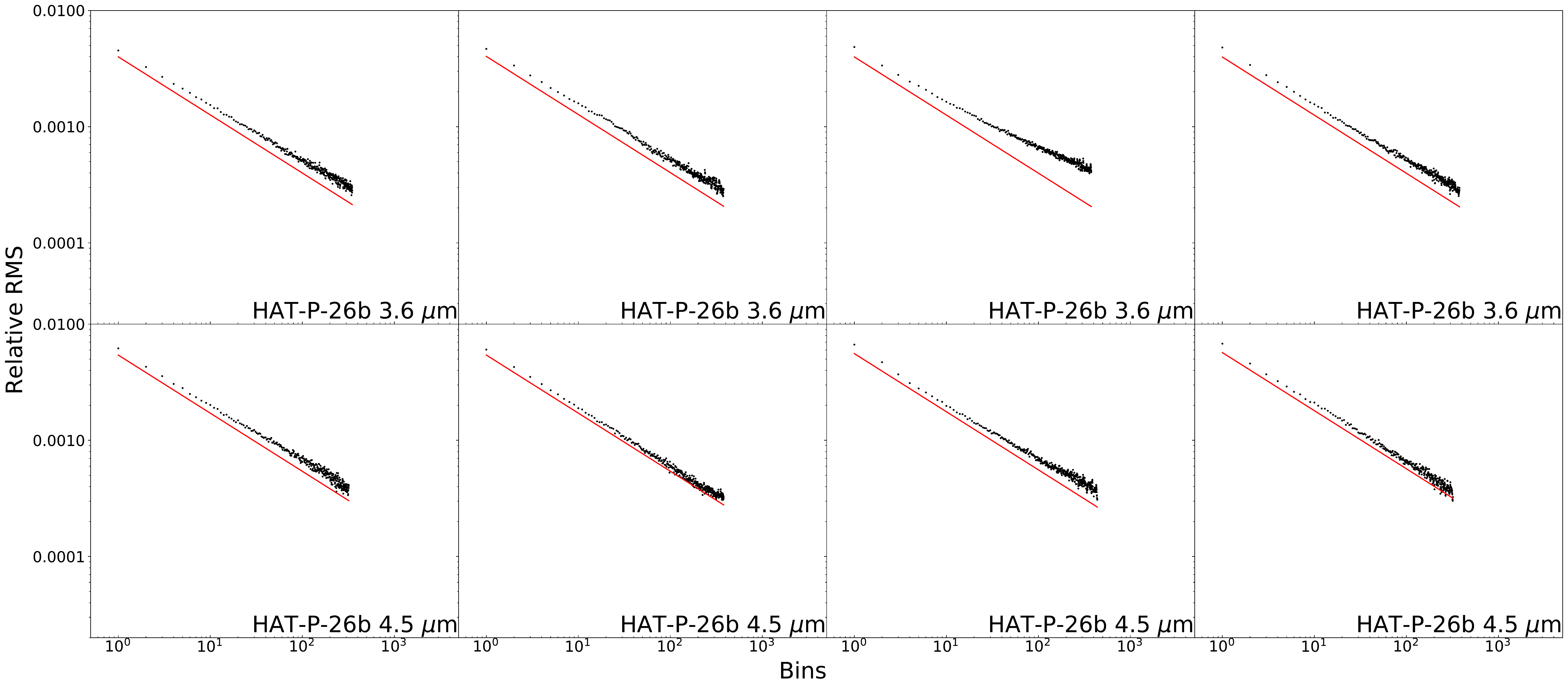}
\caption{Standard deviation of the residuals of each visit of HAT-P-26b after removing the best-fit instrumental and astrophysical models as a function of bin size. The solid lines show the photon noise limits as a function of bin size scaled by 1/$\sqrt{N}$. Observations are shown in chronological order across each row. }
\label{fig:A7}  
\end{figure*}
\end{centering}

\begin{centering}
 \begin{figure*}[h!]
\includegraphics[width=\textwidth]{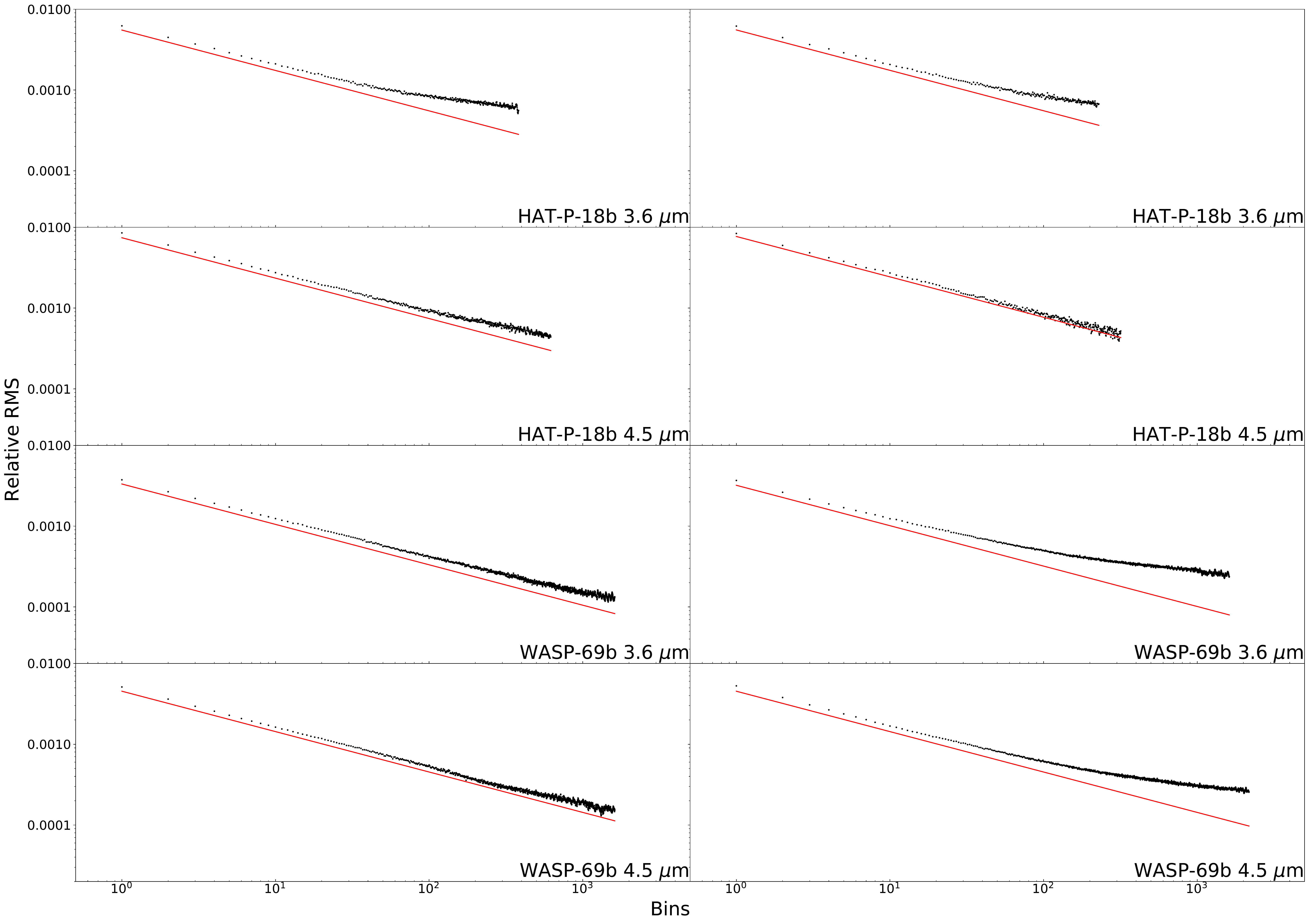}
\caption{Standard deviation of the residuals of each visit of HAT-P-18b and WASP-69b after removing the best-fit instrumental and astrophysical models as a function of bin size. The solid lines show the photon noise limits as a function of bin size scaled by 1/$\sqrt{N}$. Observations are shown in chronological order across each row. }
\label{fig:A8}  
\end{figure*}
\end{centering}

\begin{centering}
\begin{figure*}[h!]
\includegraphics[width=\textwidth]{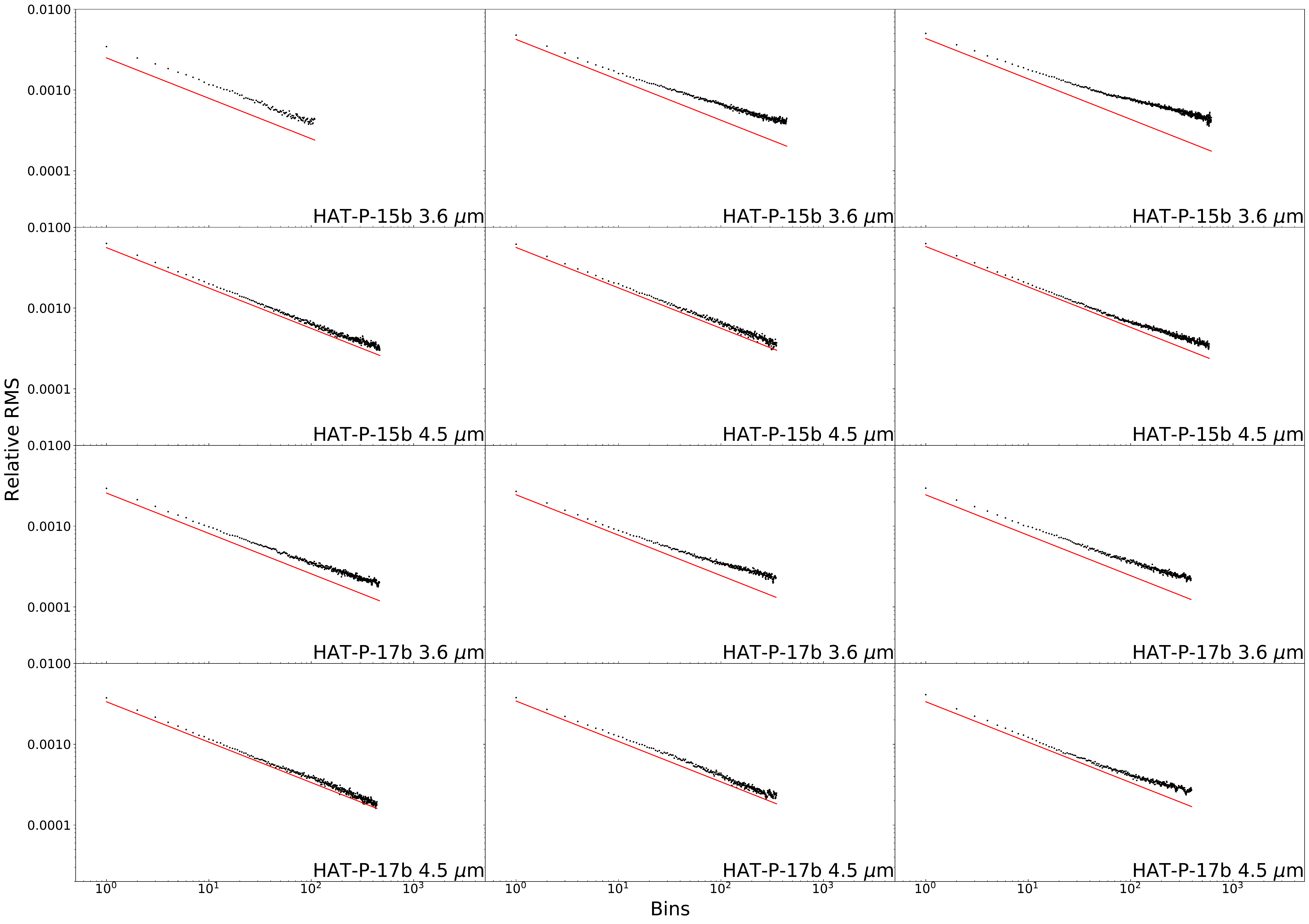}
\caption{Standard deviation of the residuals of each visit of HAT-P-15b and HAT-P-17b after removing the best-fit instrumental and astrophysical models as a function of bin size. The solid lines show the photon noise limits as a function of bin size scaled by 1/$\sqrt{N}$. Observations are shown in chronological order across each row. }
\label{fig:A9}  
\end{figure*}
\end{centering}

We show the individual normalized raw light curves for each visit of HAT-P-15b, HAT-P-17b, HAT-P-18b, HAT-P-26b, and WASP-69b in Figures ~\ref{fig:A1} -~\ref{fig:A3}. For each of the visits, we overplot the best-fit instrumental noise model derived from the joint fits. We then show these same individual light curves with the instrumental noise models removed in Figures ~\ref{fig:A4} -~\ref{fig:A6} and overplot the best-fit eclipse model derived from the joint fits. We show the standard deviation of the residuals after removing both the eclipse and noise models from each visit for each planet in Figures ~\ref{fig:A7} -~\ref{fig:A9}. The predicted photon noise for each visit is overplotted for reference. 

\begin{figure*}[h!]
\begin{centering}
\includegraphics[width=.75\textwidth]{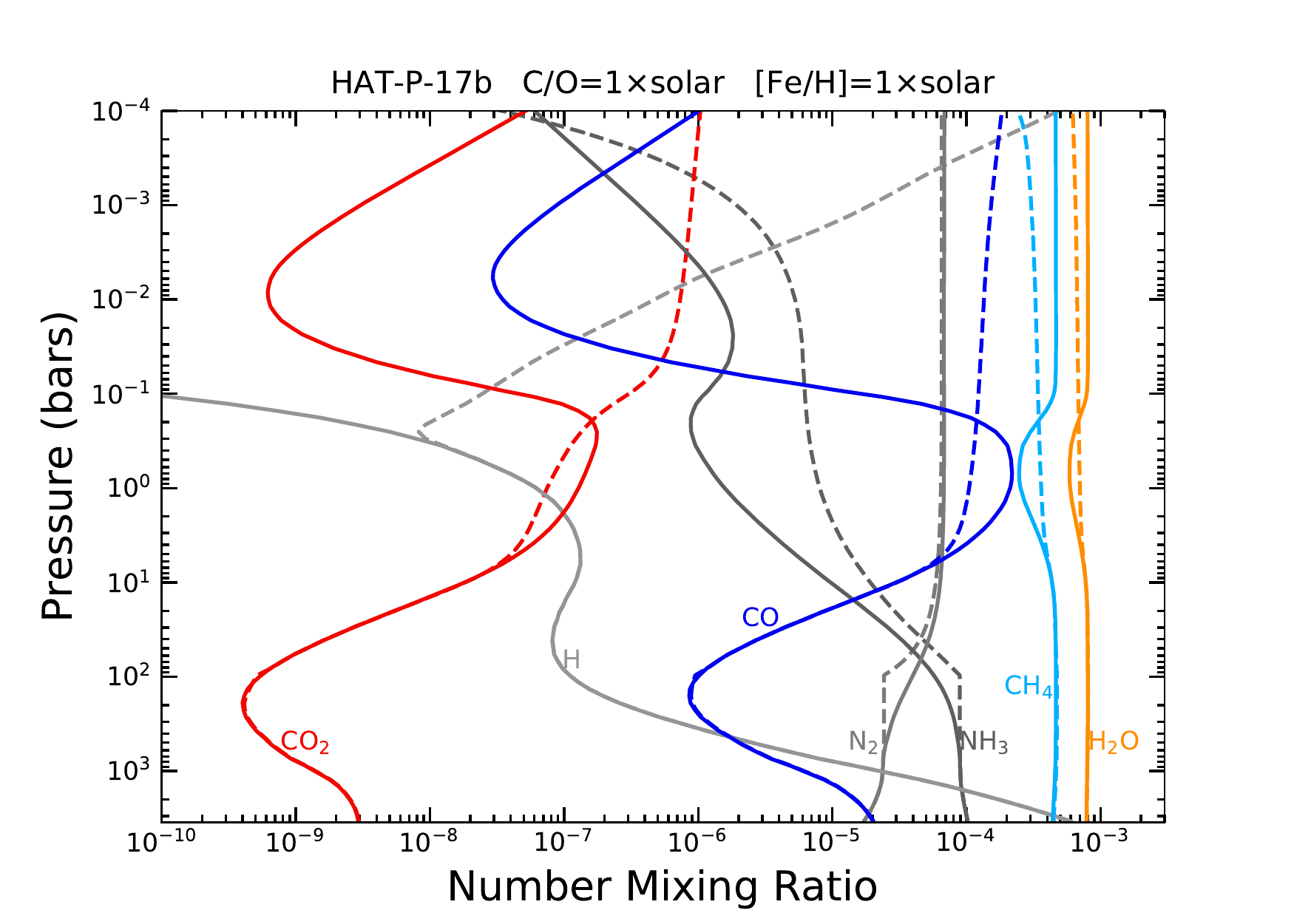}
\caption{Mixing ratio profiles derived using the same framework as presented in Moses et al. (\citeyear{moses11}, \citeyear{ Moses2013Compositional436b}, \citeyear{moses16}) for relevant species of interest for a 1$\times$ solar metallicity equilibrium (solid lines) and disequilibrium (dashed lines) model with solar C/O ratios for HAT-P-17b.}
\label{fig:h17_z1}  
\end{centering}
\end{figure*}

\begin{figure*}[h!]
\begin{centering}
\includegraphics[width=.75\textwidth]{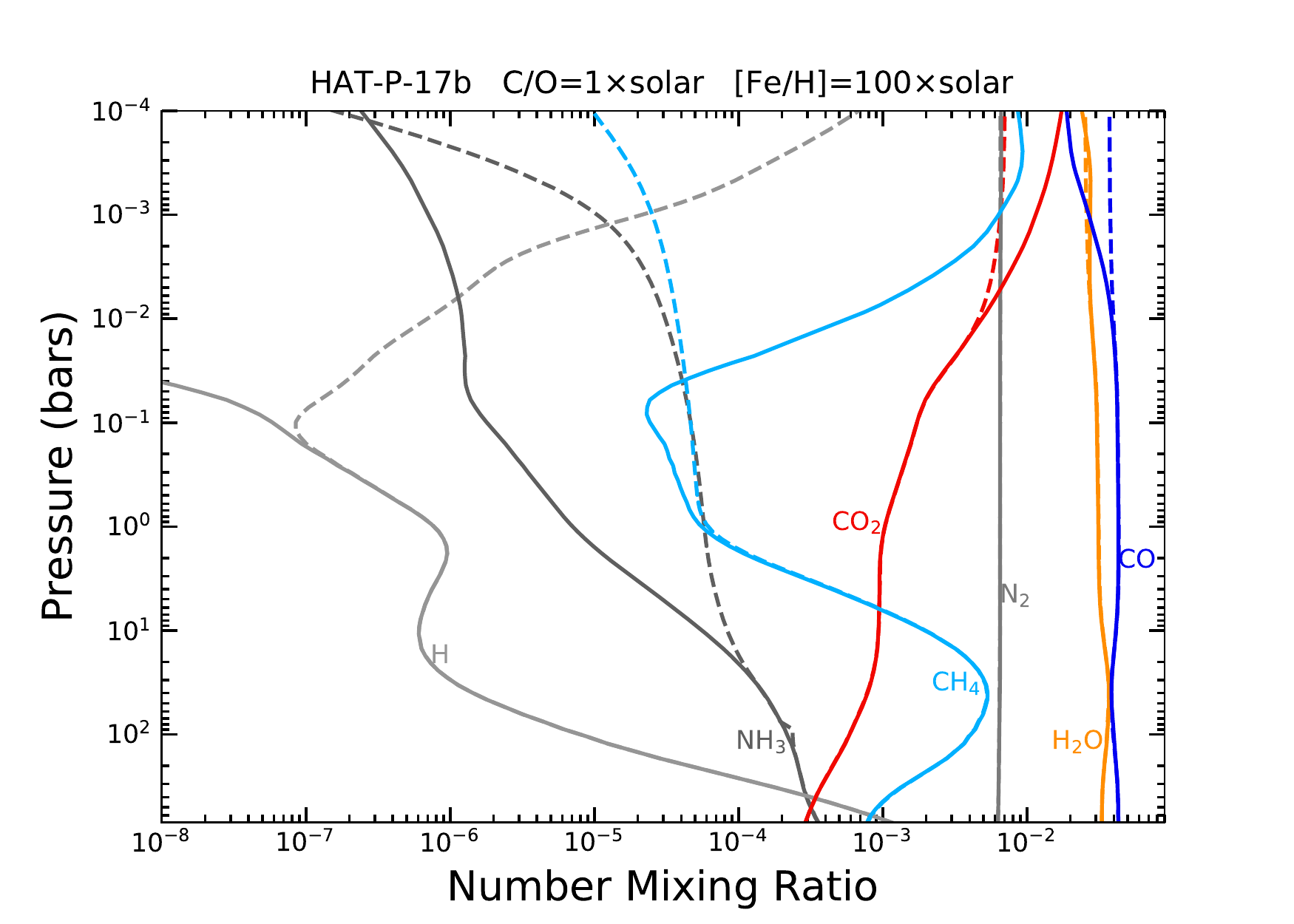}
\caption{Mixing ratio profiles derived using the same framework as presented in Moses et al. (\citeyear{moses11}, \citeyear{ Moses2013Compositional436b}, \citeyear{moses16}) for relevant species of interest for a 100$\times$ solar metallicity equilibrium (solid lines) and disequilibrium (dashed lines) model with solar C/O ratios for HAT-P-17b.}
\label{fig:h17_z100}  
\end{centering}
\end{figure*}

\begin{figure*}[h!]
\begin{centering}
\includegraphics[width=.75\textwidth]{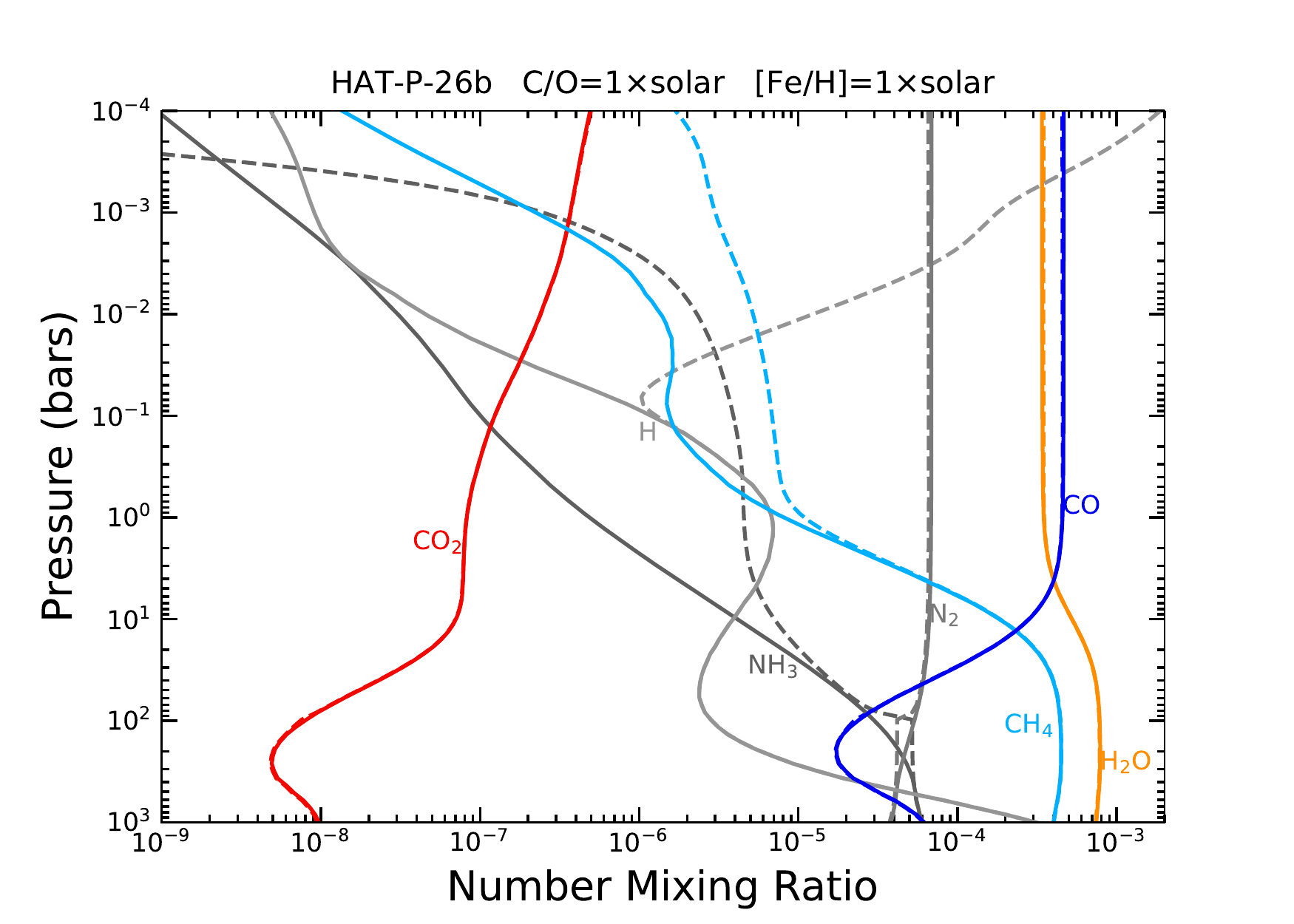}
\caption{Mixing ratio profiles derived using the same framework as presented in Moses et al. (\citeyear{moses11}, \citeyear{ Moses2013Compositional436b}, \citeyear{moses16}) for relevant species of interest for a 1$\times$ solar metallicity equilibrium (solid lines) and disequilibrium (dashed lines) model with solar C/O ratios for HAT-P-26b.}
\label{fig:h26_z1}  
\end{centering}
\end{figure*}

\begin{figure*}[h!]
\begin{centering}
\includegraphics[width=.75\textwidth]{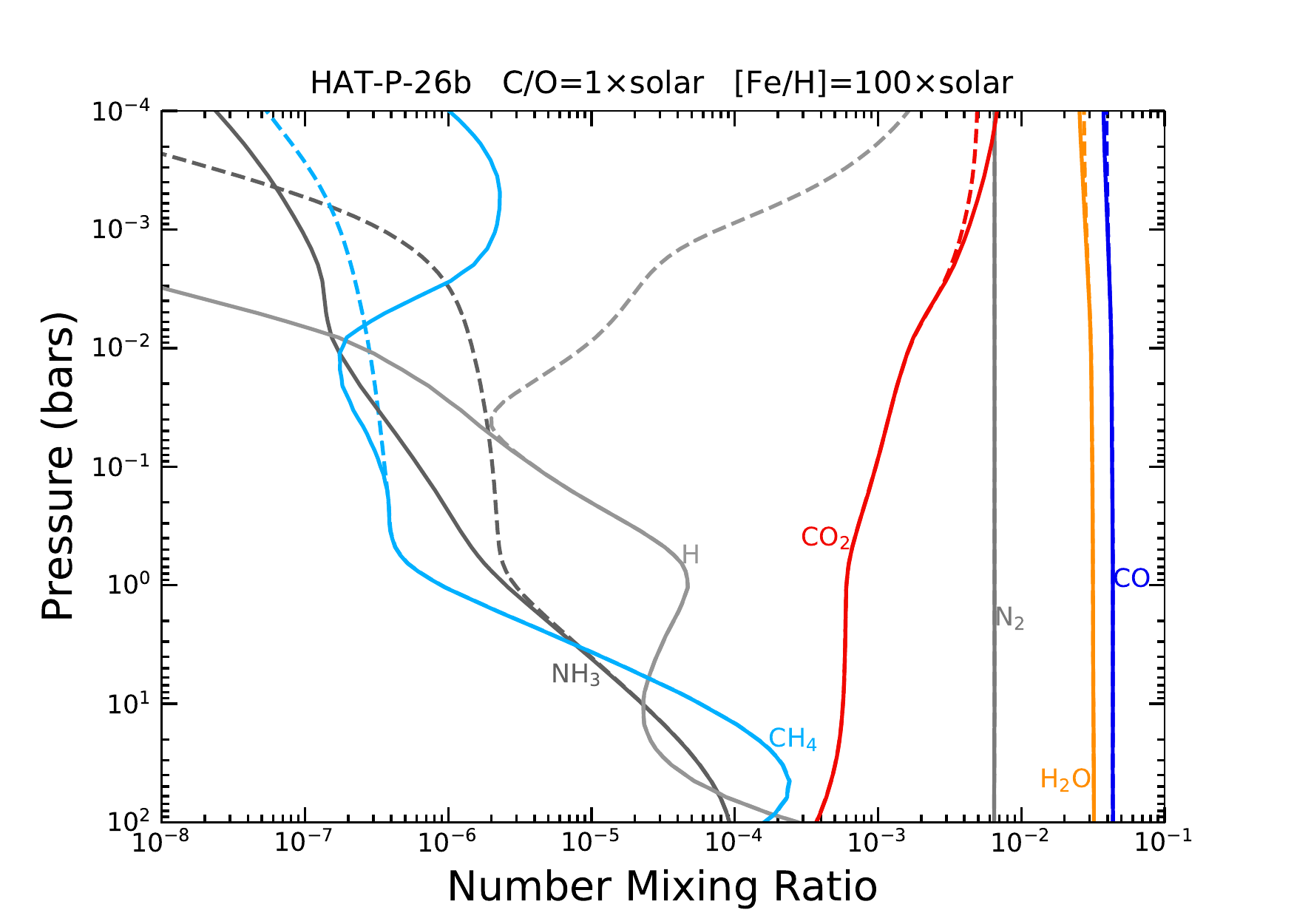}
\caption{Mixing ratio profiles derived using the same framework as presented in Moses et al. (\citeyear{moses11}, \citeyear{ Moses2013Compositional436b}, \citeyear{moses16}) for relevant species of interest for a 100$\times$ solar metallicity equilibrium (solid lines) and disequilibrium (dashed lines) model with solar C/O ratios for HAT-P-26b.}
\label{fig:h26_z100}  
\end{centering}
\end{figure*}

We show the mixing-ratio profiles for relevant species of interest for a 1$\times$ and 100$\times$ solar metallicity model for HAT-P-17b (Figures  \ref{fig:h17_z1} and  \ref{fig:h17_z100} respectively) and a 1$\times$ and 100$\times$ solar metallicity model for HAT-P-26b (Figures  \ref{fig:h26_z1} and  \ref{fig:h26_z100} respectively).  We show both equilibrium and disequilibrium chemical models derived using the same framework as presented in Moses et al. (\citeyear{moses11}, \citeyear{ Moses2013Compositional436b}, \citeyear{moses16}).
\clearpage
\counterwithin{figure}{section}
\section{Figures}

\end{document}